\documentclass[superscriptaddress, prd, aps,amsmath,amssymb,showpacs,showkeys, onecolumn]{revtex4}
\usepackage[dvips]{graphicx,color}
\usepackage{subfig}
\usepackage{times}
\usepackage{xcolor}
\usepackage[colorlinks=true,urlcolor=blue,linkcolor=red,citecolor=blue]{hyperref}

\begin{document}
\title{Photon spheres, gravitational lensing/mirroring, and greybody radiation in deformed AdS-Schwarzschild black holes with phantom global monopole}

\author{Faizuddin Ahmed}
\email[Email: ]{faizuddinahmed15@gmail.com}

\affiliation{Department of Physics, University of Science \& Technology Meghalaya, Ri-Bhoi, Meghalaya, 793101, India}

\author{Ahmad Al-Badawi}
\email[Email: ]{ahmadbadawi@ahu.edu.jo}
\affiliation{Department of Physics, Al-Hussein Bin Talal University 71111, Ma’an, Jordan}

\author{ \.{I}zzet Sakall}
\email[Email: ]{izzet.sakalli@emu.edu.tr}

\affiliation{Physics Department, Eastern Mediterranean University, Famagusta 99628, North Cyprus via Mersin 10, Turkey}

\begin{abstract}
In this study, we investigate the geodesic structure, gravitational lensing/mirroring phenomena, and scalar perturbations of deformed AdS-Schwarzschild black holes with global monopoles, incorporating both ordinary and phantom configurations. We introduce a modified black hole metric characterized by a deformation parameter $\alpha$, a control parameter $\beta$, and a symmetry-breaking scale parameter $\eta$, which collectively influence the spacetime geometry. Through comprehensive geodesic analysis, we determine the photon sphere radius numerically for various parameter configurations, revealing significant differences between ordinary and phantom global monopoles. The stability of timelike circular orbits is assessed via the Lyapunov exponent, demonstrating how these parameters affect orbital dynamics. Our gravitational lensing analysis, employing the Gauss-Bonnet theorem, reveals a remarkable gravitational mirroring effect in phantom monopole spacetimes at high AdS curvature radii, where light rays experience negative deflection angles-being repelled rather than attracted by the gravitational field. Furthermore, we analyze massless scalar perturbations and derive the corresponding greybody factors, which characterize the transmission of Hawking radiation through the effective potential barrier surrounding the black hole. Our numerical results indicate that phantom global monopoles substantially modify both gravitational lensing/mirroring properties and the radiation spectrum compared to ordinary monopoles. The presence of the deformation parameter $\alpha$ introduces additional complexity to the system, leading to distinct thermodynamic behavior that deviates significantly from the standard AdS-Schwarzschild solution.
\end{abstract}

\keywords{Corrected entropy; Deformed black hole; Black hole thermodynamics; Black hole phase transition; Gravitational lensing; Phantom global monopole.}

\maketitle

\section{Introduction}\label{sec:1}

The study of black holes (BHs) within anti-de Sitter (AdS) spacetime represents one of the most profound developments in gravitational physics. The AdS-Schwarzschild BH solution has garnered substantial attention due to its pivotal role in the AdS/CFT correspondence, which establishes a duality between gravitational systems in asymptotically AdS spacetimes and conformal field theories in one fewer dimension \cite{isz01,isz02,isz03}. This correspondence provides a powerful framework for investigating strongly coupled quantum systems, offering profound insights into quantum gravity and high-energy physics \cite{isz04,isz05}.

Deformed AdS-Schwarzschild BHs represent significant extensions of the standard solution, incorporating additional gravitational degrees of freedom that may emerge from quantum gravitational effects or alternative gravitational frameworks \cite{isz06,isz07}. These deformations modify the BH metric while preserving its asymptotic AdS behavior, leading to distinctive gravitational and thermodynamic properties \cite{isz08,isz09}. The deformation parameter $\alpha$ introduces corrections to the metric function that systematically alter the spacetime structure, potentially revealing signatures of physics beyond general relativity (GR). The incorporation of a control parameter $\beta$ further enriches the solution by regulating behavior near the central singularity, effectively modeling quantum gravitational effects that may ameliorate classical singularities \cite{isz12}.

Global monopoles (GMs) constitute topological defects that emerge during spontaneous symmetry breaking in the early universe, particularly when a global $O(3)$ symmetry is broken to $O(2)$ \cite{isz13,isz14}. These defects manifest as point-like sources that modify the surrounding spacetime geometry, generating a solid deficit angle and altering the asymptotic structure \cite{isz15}. The gravitational influence of GMs extends to BH physics, where their presence modifies the metric structure, leading to significant deviations from standard solutions \cite{isz16,isz17}. Of particular interest are phantom GMs, characterized by negative energy density contributions that violate the null energy condition, potentially yielding repulsive gravitational effects \cite{isz18,isz19}. This phantom behavior arises from exotic matter fields that play critical roles in various cosmological scenarios, including dark energy models and wormhole configurations \cite{isz20,isz21}.

The investigation of geodesic structure in BH spacetimes provides essential insights into their gravitational properties and observational signatures. Null geodesics, which describe the trajectories of massless particles such as photons, are especially significant as they determine the optical appearance of BHs and their lensing characteristics \cite{isz22,isz23}. The photon sphere, a critical surface where photons can orbit the BH in unstable circular paths, establishes a fundamental length scale that influences numerous astrophysical phenomena, including BH shadows and gravitational lensing \cite{isz24,isz25}. For standard Schwarzschild BHs, the photon sphere occurs at $r = 3M$, but this radius is modified in the presence of deformations and GMs, leading to distinctive observational signatures \cite{isz26}. The effective potential governing geodesic motion encapsulates the gravitational influence of the BH, with its structure determining critical features such as the innermost stable circular orbit (ISCO) for massive particles and the unstable circular orbit for photons \cite{isz27,isz28}.

Gravitational lensing represents one of the most powerful observational probes of BH spacetimes, providing direct access to the gravitational field structure \cite{isz29,isz30}. The deflection of light by gravitational sources constitutes a fundamental prediction of GR that has been confirmed through numerous observations \cite{isz31}. In the context of BH physics, the Gauss-Bonnet theorem (GBTh) offers an elegant topological approach to calculating the deflection angle in the weak-field limit, relating it to the integral of the Gaussian curvature over a suitable domain \cite{isz32,isz33}. This approach reveals how spacetime deformations and GMs modify the deflection angle, potentially yielding observable deviations from standard GR predictions \cite{isz34}. These modifications are particularly relevant for current and future observational missions, including the Event Horizon Telescope, which aims to resolve the shadow of supermassive BHs with unprecedented precision \cite{isz35}. A few recent works on gravitational lensing in various curved spacetimes were reported in \cite{aa1,aa2,aa3}.

A particularly intriguing phenomenon that emerges in our analysis is the gravitational mirroring effect associated with phantom GMs in high AdS curvature regimes \cite{isz101,isz102,isz103,isz104,isz105,isz106}. Unlike conventional gravitational lensing, where light rays are invariably bent toward the gravitating source, phantom GMs can induce negative deflection angles at certain impact parameters, effectively repelling rather than attracting light rays. This counterintuitive behavior stems from the violation of energy conditions characteristic of phantom matter, which introduces effective negative gravitational mass contributions that fundamentally alter the optical properties of the spacetime. The gravitational mirroring effect represents a distinctive observational signature that could potentially distinguish phantom topological defects from other exotic matter distributions, providing a crucial test for modified gravitational theories that accommodate energy condition violations. Such phenomena have been theoretically explored in related contexts involving negative energy densities, though the specific manifestation in deformed AdS backgrounds with phantom GMs presents novel features with significant observational implications.

On the other hand, the thermal aspects of BH physics, encapsulated in Hawking radiation and its spectral characteristics, provide a fundamental connection between gravitational physics, thermodynamics, and quantum field theory \cite{isz36,isz37}. While Hawking's original calculation predicted a perfect blackbody spectrum, the actual emission from BHs is modified by the gravitational potential surrounding the horizon, leading to the concept of Greybody Factors (GFs) \cite{isz38}. These factors represent the transmission probability for radiation to escape from the BH horizon to spatial infinity, effectively encoding how the background spacetime filters the emitted radiation \cite{isz39}. The calculation of GFs involves solving the wave equation for perturbations propagating in the BH background, with the resulting transmission coefficient depending on the radiation frequency and BH properties \cite{isz40}. For deformed AdS-Schwarzschild BHs with GMs, these GFs exhibit distinctive features that reflect the modified spacetime structure, potentially providing observational signatures of exotic physics.

In Ref. \cite{isz41}, the authors derived a deformed AdS-Schwarzschild BH solution using the gravitational decoupling method, incorporating an additional gravitational source that satisfies the weak energy condition. They had chosen a monotonic energy density function consistent with these constraints. Their method involves a positive deformation parameter that adjusts the strength of geometric deformations on the background geometry. They considered a four-dimensional action with an additional general Lagrangian term to modify standard Einstein gravity with a cosmological constant.
\begin{equation}
    S_\text{deformed AdS-Schwarzschild BH}=\int \sqrt{-g}\,d^4x \left(\frac{R}{2}+\frac{3}{\ell^2_p}+L_m+L_X\right),\label{action1}
\end{equation}
where $g$ is the determinant of the metric, $R$ is the Ricci scalar, $\ell_p$ is the curvature radius related with the cosmological constant as $\frac{1}{\ell^2_p}=-\frac{\Lambda}{3}$, $L_m$ is the usual matter Lagrangian and $L_X$ represents Lagrangian for any other matter or new gravitational sector beyond general relativity, {\it e.g.}, Lovelock gravity and/or new other scalar/vector/tensor field(s). Through a rigorous treatment, the authors obtained the following metric function
\begin{equation}
    f(r)=1-\frac{2\,M}{r}+\frac{r^2}{\ell^2_p}+\frac{\alpha\,(\beta^2+3\,r^2+3\,r\,\beta)}{3\,r\,(r+\beta)^2},\label{function}
\end{equation}
by considering the following spherically symmetric metric
\begin{equation}
    ds^2=-f(r)\,dt^2+\frac{dr^2}{f(r)}+r^2\,(d\theta^2+\sin^2 \theta\,d\phi^2).\label{function2}
\end{equation}

In Ref. \cite{SCJJ}, the authors constructed a spherically symmetric BH with phantom global monopole. The line-element is described by
\begin{equation}
ds^2=-\mathcal{A}(r)\,dt^2+\frac{dr^2}{\mathcal{A}(r)}+r^2\,(d\theta^2+\sin^2 \theta\,d\phi^2), \label{a1}
\end{equation}
with the metric function
\begin{equation}
    \mathcal{A}(r)=\Big(1-8\,\pi\,\eta^2\,\xi-\frac{2\,M}{r}+\frac{r^2}{\ell^2_p}\Big),\label{a2}
\end{equation}
where $\eta$ and $\xi$ stand for the energy scale of symmetry breaking, and BH kinetic energy, respectively. The scenario mediated by $\xi=1$ describes an arena of an ordinary global monopole that arises from the scalar field’s non-negative and non-zero kinetic energy \cite{isz13}. However, the phantom global monopole is generated by selecting $-1$ value of $\xi$, thereby relating it with the scalar field’s negative kinetic energy. The above metric was obtained after a rigorous analysis using the following action given by \cite{SCJJ}
\begin{equation}
    S_\text{phantom global monopole}=\int \sqrt{-g}\,d^4x\,\left[\frac{R}{2}+\frac{3}{\ell^2_p}-\frac{\xi}{2}\,(\partial^{\mu}\psi^i)\,(\partial_{\mu}\psi^i)-\frac{\lambda}{4}\,(\psi^i\,\psi^i-\eta^2)^2\right].\label{phantom}
\end{equation}
Here $\psi^i$ represents the triplet scalar field with $i=1,2,3$, $\eta$ is the energy scale of symmetry breaking and $\lambda$ is a constant. The coupling constant $\xi$ in the kinetic term takes the value $\xi=1$ corresponds to the case of an ordinary global monopole originating from the scalar field with the positive kinetic energy \cite{isz13}. As the coupling constant $\xi=-1$, the kinetic energy of the scalar field is negative and then the phantom global monopole is formed.

Our investigation is motivated by the recognition that the interplay between spacetime deformations and topological defects yields rich physical phenomena extending beyond standard GR predictions. By incorporating both deformation parameters and GMs into the AdS-Schwarzschild solution, we explore how these modifications collectively influence fundamental aspects of BH physics, including geodesic structure, gravitational lensing/mirroring, and Hawking radiation. The introduction of phantom GMs adds a particularly intriguing dimension, as their exotic matter content violates standard energy conditions, potentially leading to repulsive gravitational effects that significantly alter both geodesic motion and radiation characteristics. Furthermore, the AdS asymptotic structure provides a natural framework for exploring these phenomena in holographic contexts, where boundary CFT interpretations may offer additional insights. The interplay between nonlinear electrodynamics (NLED) and modified BH solutions has attracted significant attention, as NLED introduces corrections to the electromagnetic sector that can substantially alter BH properties \cite{isz22,isz23}. While our present investigation focuses on neutral BH configurations, the methodologies developed here establish a foundation for future studies incorporating NLED effects in deformed AdS backgrounds with topological defects. In short, our primary objectives in this study are to: (1) establish the metric structure of deformed AdS-Schwarzschild BHs with both ordinary and phantom GMs, analyzing how the parameters $\alpha$, $\beta$, and $\eta$ shape the spacetime geometry; (2) conduct a comprehensive analysis of geodesic motion, with emphasis on determining the photon sphere radius and its parameter dependence; (3) investigate circular orbit stability through Lyapunov exponent calculations; (4) apply the GBTh to derive the gravitational deflection angle in the weak-field limit, characterizing both conventional lensing and exotic mirroring effects; and (5) analyze scalar perturbations, calculating the GFs that characterize Hawking radiation transmission.

The paper is organized as follows. In Section \ref{sec:2}, we introduce the deformed AdS-Schwarzschild BH with phantom GM, establishing the metric structure and fundamental properties. There, an analysis of geodesic motion of test particles, including photon spheres is presented, and a study of the impact of phantom GM on photon and timelike particles. In Section \ref{sec:3}, we analyze gravitational lensing and mirroring phenomena through deflection angle calculations using the GBTh, paying particular attention to the emergence of negative deflection angles in phantom GM spacetimes. Section \ref{sec:4} focuses on scalar perturbations, deriving the wave equation and analyzing the modified potential, including GF calculations. Finally, in Section \ref{sec:5}, we summarize our findings and discuss implications for BH physics and modified gravity theories.

\section{Static and spherically symmetric deformed AdS-Schwarzschild BH: Geodesics analysis, scalar perturbations} \label{sec:2}

In this section, we present a detailed analysis of the deformed AdS-Schwarzschild BH metric modified by the presence of GMs, focusing on the resulting geodesic structure and perturbative dynamics. The spacetime geometry under consideration builds upon prior investigations of deformed BH solutions \cite{isz41} and BH solution with phantom GM \cite{SCJJ}, which significantly alter the effective gravitational field.

Therefore, we begin by considering a static, spherically symmetric deformed AdS-Schwarzschild BH with phantom GM, characterized by the following line-element in natural units:
\begin{equation}
   ds^2=-\mathcal{F}(r)\,dt^2+\frac{dr^2}{\mathcal{F}(r)}+r^2\,(d\theta^2+\sin^2 \theta\,d\phi^2),\label{bb1}
\end{equation}
where the metric function $\mathcal{F}(r)$ incorporates both the deformation parameters and the phantom GM contribution:
\begin{equation}
   \mathcal{F}(r)=1-8\,\pi\,\eta^2\,\xi-\frac{2\,M}{r}+\frac{r^2}{\ell^2_{p}}+\frac{\alpha \beta^2}{3\,r\,(\beta+r)^3}+\frac{\alpha}{(\beta+r)^2}.\label{bb2}
\end{equation}
Here $\alpha$ is the deformation parameter with dimensions of length squared, $\beta$ is the control parameter with dimensions of length that regulates behavior near the central singularity. In the limit where $\xi=0$, the metric function (\ref{bb2}) with spacetime (\ref{bb1}) reduces to the deformed AdS-Schwarzschild BH solution, reported in Ref. \cite{isz41}. Moreover, in the limit where $\xi=0$, and $\beta=0$, the metric function (\ref{bb2}) becomes $\mathcal{F}(r)=1-\frac{2\,M}{r}+\frac{r^2}{\ell^2_{p}}+\frac{\alpha}{r^2}$, and hence the spacetime reduces to an AdS BH with quantum-like potential, reported in \cite{AFA}. In addition, in the limit where $\alpha=0$, the metric function becomes $\mathcal{F}(r)=1-8\,\pi\,\eta^2\,\xi-\frac{2\,M}{r}+\frac{r^2}{\ell^2_{p}}$, and hence the spacetime (\ref{bb1}) reduces to an AdS BH solution with phantom global monopole, reported in \cite{SCJJ} which reduces further to AdS-Schwarzschild BH when $\xi=0$. 

For the above selected BH spacetime, we determine a few scalar quantities. The Ricci scalar $R$ is given by
\begin{equation}
    R=g^{\mu\nu}\,R_{\mu\nu}=4\,\left[\frac{\alpha\,\beta}{(\beta+r)^2}+\Lambda+\frac{4\,\pi\,\eta^2\,\xi}{r^2} \right].\label{ricci}
\end{equation}
The Kretschmann scalar $\mathcal{K}=R^{\mu\nu\lambda\sigma}\,R_{\mu\nu\lambda\sigma}$ is given by
\begin{eqnarray}
&&\mathcal{K}=\frac{4}{3\,r^6\,(\beta+r)^{10}}\,\Big[42\,r^8\, \alpha^2 + 120\, r^7\, \alpha1^2\, \beta+ 
 222\, r^6\, \alpha^2\, \beta^2 + 260\, r^5\, \alpha^2\, \beta^3+212\, r^4\, \alpha^2\, \beta^4 + 120\, r^3\,\alpha^2\, \beta^5+45\, r^2\, \alpha^2\, \beta^6\nonumber\\
 && +10\,r\, \alpha^2\, \beta^7 + \alpha^2\, \beta^8+36\, M^2\, (r + \beta)^{10} + 
 4\, r^{11}\, \alpha\,\beta\,\Lambda+ 
 20\, r^{10}\, \alpha\,\beta^2\, \Lambda+40\, r^9\, \alpha\,\beta^3\, \Lambda+ 
 40\, r^8\, \alpha\,\beta^4\, \Lambda + 
 20\, r^7\, \alpha\,\beta^5\, \Lambda\nonumber\\
 &&+ 
 4\, r^6\, \alpha\,\beta^6\, \Lambda+ 
 2\, r^{16}\, \Lambda^2 + 20\,r^{15}\, \beta\,\Lambda^2 + 
 90\, r^{14}\, \beta^2\, \Lambda^2+240\, r^{13}\, \beta^3\,\Lambda^2 + 
 420\, r^{12}\, \beta^4\, \Lambda^2 + 
 504\, r^{11}\,\beta^5\,\Lambda^2 + 
 420\,r^{10}\,\beta^6\,\Lambda^2 \nonumber\\
 &&+ 
 240\, r^9\,\beta^7\, \Lambda^2 + 
 90\,r^8\,\beta^8\,\Lambda^2
 +20\, r^7\,\beta^9\,\Lambda^2 + 
 2\, r^6\,\beta^{10}\,\Lambda^2 +16\,\pi\,r\,(r + \beta)^7\,\eta^2\,\left\{-\alpha\,(3\,r^2 + 3\, r\, \beta+\beta^2)+r^3\, (r + \beta)^3\,\Lambda\right\}\,\xi\nonumber\\
 &&+192\,\pi^2\, r^2\, (r + \beta)^{10}\,\eta^4\,\xi^2
 +12\, M\, (r + \beta)^5\,\left\{-\alpha\,(6\,r^4 + 10\, r^3\, \beta+10\,r^2\,\beta^2 + 5\, r\,\beta^3 + \beta^4)+ 
    8\, \pi\,r\,(r + \beta)^5\,\eta^2\,\xi\right\}\Big].\label{ricci2}
\end{eqnarray}
It is evident that these scalar quantities diverge at $r=0$, specifically,
\begin{equation}
    \lim_{r \to 0}\,R =\infty,\quad\quad \lim_{r \to 0}\,\mathcal{K} =\infty\label{ricci3}
\end{equation}
indicating the presence of a curvature or central singularity at the origin. Furthermore, at large distances, $r \to  \infty$, we have 
\begin{equation}
    \lim_{r \to \infty}\,R = 4\,\Lambda,\quad\quad \lim_{r \to \infty}\,\mathcal{K} =\frac{8\,\Lambda^2}{3}\label{ricci4}
\end{equation}
suggesting that the selected spacetime asymptotically approaches the structure of an Anti-de Sitter (AdS) space at large distances. These observations confirm that the deformed BH metric (\ref{bb1}) under consideration represents a singular BH solution in general relativity.

Taking the radial derivative of Eq.~(\ref{bb2}) and multiplying by $r$ yields:
\begin{equation}
   r\,\mathcal{F}'(r)=\frac{2\,M}{r}+\frac{2\,r^2}{\ell^2_{p}}-\left(\frac{\alpha\,\beta^2}{r\,(\beta+r)^3}+\frac{3\,\alpha\,\beta^2}{(\beta+r)^4}\right)-\frac{2\,\alpha\,r}{(\beta+r)^3},\label{bb3}
\end{equation}
where prime denotes differentiation with respect to $r$. From Eqs.~(\ref{bb2}) and (\ref{bb3}), we can compute the quantity $(2\,\mathcal{F}(r)-r\,\mathcal{F}'(r))$, which plays a critical role in determining the photon sphere:
\begin{equation}
   2\,\mathcal{F}(r)-r\,\mathcal{F}'(r)=2\,\left[1-8\,\pi\,\eta^2\,\xi-\frac{3\,M}{r}+\frac{\alpha}{(\beta+r)^2}+\frac{\alpha\,\beta^2}{2\,r\,(\beta+r)^3}+\frac{\alpha\,r}{(\beta+r)^3}+\frac{\alpha\,\beta^2}{2\,(\beta+r)^4}\right].\label{bb4}
\end{equation}

The explicit equation governing the photon sphere radius, derived from the condition $r\mathcal{F}'(r) = 2\mathcal{F}(r)$, is given by:
\begin{equation}
\left(2\,r^{2}+4\,\alpha +8\,r\,\beta \right)\, r^{3}+\left( 6\,r^{2}+4\,r\,\beta +\beta^{2}\right)\, \alpha\, \beta +\left( 6\,r^{2}+4\,r\,\beta +\beta ^{2}\right)\, 2\,r\,\beta^{2}-\left( 6\,M+2\,r\,\eta^{2}\,\xi \right) \left( r+\beta \right) ^{4}=0.\label{rpha1}
\end{equation}

This equation exhibits high nonlinearity and cannot be solved analytically for the photon sphere radius $r_{ph}$. We therefore employ numerical methods to determine $r_{ph}$ for various parameter configurations. Tables \ref{table1a} and \ref{table2a} present these numerical results, revealing the contrasting behaviors of photon spheres in spacetimes with an ordinary versus a phantom GM.

\begin{center}
\begin{tabular}{|c|c|c|c|c|c|c|c|c|c|c|c|c|c|c|c|}
 \hline \multicolumn{13}{|c|}{Photon sphere $r_{ph}$ for ordinary GMs }
 \\ \hline
  $\xi=1$ &\multicolumn{4}{|c|}{ $\eta=0.2$}&\multicolumn{4}{|c|}{ $\eta=0.4$}&\multicolumn{4}{|c|}{ $\eta=0.6$} \\ \hline 
$\alpha $ & $\beta =0.2$ & $0.4$ & $0.6$ & $0.8$ & $0.2$ & $0.4$ & $%
0.6$ & $0.8$ & $0.2$ & $0.4$ & $0.6$ & $0.8$ \\ \hline
$0.2$ & $3.00672$ & $3.023$ & $3.03577$ & $3.04602$ & $3.45134$ & $3.46609$
& $3.47799$ & $3.48777$ & $4.56435$ & $4.57625$ & $4.58634$ & $4.59497$ \\ 
$0.4$ & $2.87964$ & $2.9156$ & $2.94309$ & $2.96474$ & $3.32328$ & $3.35555$
& $3.38105$ & $3.40167$ & $4.43476$ & $4.46036$ & $4.48176$ & $4.49988$ \\ 
$0.6$ & $2.74145$ & $2.80186$ & $2.84651$ & $2.88095$ & $3.18539$ & $3.23897$
& $3.28019$ & $3.31293$ & $4.2976$ & $4.33919$ & $4.3734$ & $4.40201$ \\ 
$0.8$ & $2.58861$ & $2.68052$ & $2.74554$ & $2.79443$ & $3.03502$ & $3.11524$
& $3.1749$ & $3.22129$ & $4.15138$ & $4.21196$ & $4.26082$ & $4.30112$ \\
 \hline
\end{tabular}
\captionof{table}{The photon sphere has been tabulated numerically for  deformed AdS-Schwarzschild BH with ordinary GMs for different values of BH parameters.} \label{table1a}
\end{center}

\begin{center}
\begin{tabular}{|c|c|c|c|c|c|c|c|c|c|c|c|c|c|c|c|}
  \hline \multicolumn{13}{|c|}{Photon sphere $r_{ph}$ for phantom GMs }
 \\ \hline 
  $\xi=-1$ &\multicolumn{4}{|c|}{ $\eta=0.2$}&\multicolumn{4}{|c|}{ $\eta=0.4$}&\multicolumn{4}{|c|}{ $\eta=0.6$} \\ \hline 
$\alpha $ & $\beta =0.2$ & $0.4$ & $0.6$ & $0.8$ & $0.2$ & $0.4$ & $%
0.6$ & $0.8$ & $0.2$ & $0.4$ & $0.6$ & $0.8$ \\ \hline
$0.2$ & $2.76753$ & $2.78475$ & $2.79803$ & $2.80853$ & $2.47089$ & $2.48944$
& $2.50336$ & $2.51414$ & $2.09346$ & $2.11397$ & $2.12872$ & $2.13977$ \\ 
$0.4$ & $2.64114$ & $2.67941$ & $2.70804$ & $2.73023$ & $2.34557$ & $2.38712$
& $2.41723$ & $2.44001$ & $1.97004$ & $2.0165$ & $2.0485$ & $2.07185$ \\ 
$0.6$ & $2.5028$ & $2.56759$ & $2.61423$ & $2.64954$ & $2.2071$ & $2.27818$
& $2.32739$ & $2.36366$ & $1.8316$ & $1.91239$ & $1.96486$ & $2.00201$ \\ 
$0.8$ & $2.34833$ & $2.44794$ & $2.51607$ & $2.56623$ & $2.05023$ & $2.16121$
& $2.23335$ & $2.28489$ & $1.67089$ & $1.80011$ & $1.87735$ & $1.93007$ \\
 \hline
\end{tabular}
\captionof{table}{The photon sphere has been tabulated numerically for  deformed AdS-Schwarzschild BH with Phantom GMs for different values of BH parameters.} \label{table2a}
\end{center}

Our numerical analysis in Table \ref{table1a} demonstrates that for ordinary GMs ($\xi=1$), the photon sphere radius $r_{ph}$ exhibits a systematic increase with increasing values of the symmetry-breaking parameter $\eta$. For instance, at fixed values $\alpha=0.4$ and $\beta=0.4$, the photon sphere radius increases from $r_{ph}=2.9156$ at $\eta=0.2$ to $r_{ph}=3.35555$ at $\eta=0.4$, and finally to $r_{ph}=4.46036$ at $\eta=0.6$. This represents a substantial expansion of approximately $53\%$ across this parameter range, indicating that ordinary GMs strengthen the gravitational field and push the photon sphere outward. This behavior is physically consistent with the positive energy contribution from ordinary GMs, which effectively increases the gravitational mass of the system.

In stark contrast, Table \ref{table2a} reveals that for phantom GMs ($\xi=-1$), the photon sphere radius $r_{ph}$ decreases with increasing values of $\eta$. Following the same parameter values $\alpha=0.4$ and $\beta=0.4$, we observe that $r_{ph}$ decreases from $2.67941$ at $\eta=0.2$ to $2.38712$ at $\eta=0.4$, and further contracts to $2.0165$ at $\eta=0.6$. This represents a reduction of approximately $25\%$ across the same parameter range, demonstrating that phantom GMs weaken the effective gravitational field and pull the photon sphere inward. This behavior aligns with the negative energy contribution of phantom GMs, which effectively reduces the gravitational mass of the system.

\begin{figure}
    \centering
    \includegraphics[width=0.65\linewidth]{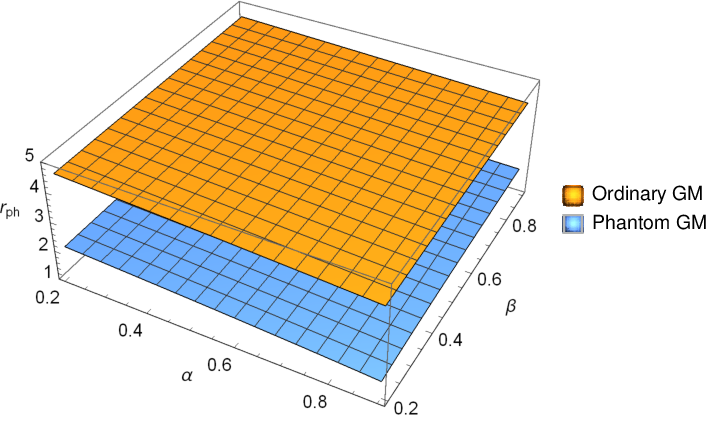}
    \caption{The profile of the photon sphere radius for various values of BH parameters $\alpha$  and $\beta$ for fixed $\eta=0.6$ showing that the $r_{ph}$ has higher value for ordinary  GMs than phantom GMs. Here, $M=1$.}
    \label{figa1}
\end{figure}
\begin{figure}[ht!]
    \centering
    \includegraphics[width=0.45\linewidth]{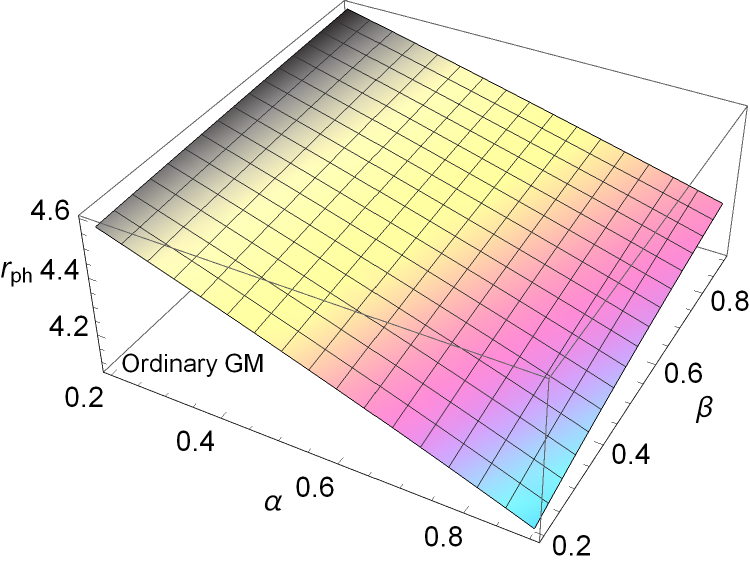}\quad\quad
    \includegraphics[width=0.45\linewidth]{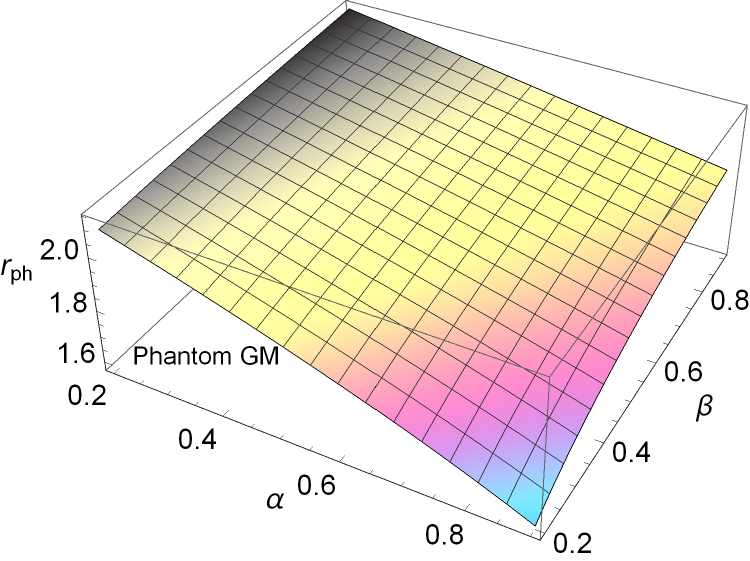}
    \caption{The profile of the photon sphere radius for various values of BH parameters $\alpha$  and $\beta$ for fixed $\eta=0.6$ showing that the $r_{ph}$ decreases with $\alpha$ but increases with $\beta$. Here, $M=1$.}
    \label{figa1a}
\end{figure}
The three-dimensional visualizations in Figures \ref{figa1} and \ref{figa1a} provide a comprehensive illustration of these contrasting behaviors. Figure \ref{figa1} clearly demonstrates the separation between the photon sphere surfaces for ordinary versus phantom GMs throughout the parameter space of $\alpha$ and $\beta$ values. The upper surface, corresponding to ordinary GMs, consistently exhibits larger photon sphere radii compared to the lower surface representing phantom GMs. This topological separation persists throughout the parameter space, confirming the fundamental difference in gravitational effects between these two classes of topological defects. Figure \ref{figa1a} displays the profile of the photon sphere radius for various values of BH parameters $\alpha$ and $\beta$ with fixed $\eta=0.6$. The $r_{ph}$ drops with $\alpha$ but grows with $\beta$.

Furthermore, both tables reveal that for fixed values of $\eta$ and $\beta$, increasing the deformation parameter $\alpha$ results in a systematic decrease in $r_{ph}$. This trend indicates that stronger metric deformations compress the effective gravitational potential, drawing the photon sphere closer to the BH horizon. Physically, this can be interpreted as the deformation parameter that introduces additional curvature contributions that enhance the gravitational binding near the BH. Conversely, for fixed values of $\alpha$ and $\eta$, increasing the control parameter $\beta$ results in a modest expansion of the photon sphere radius, suggesting that the regularization effect near the singularity slightly weakens the gravitational confinement of null geodesics.

\subsection{Geodesics Motions of Test Particles} \label{sec:2.1}

To comprehensively analyze particle dynamics in the deformed AdS-Schwarzschild BH spacetime with GMs, we investigate both null and timelike geodesics using the Lagrangian formalism. The Lagrangian density function for a test particle in a curved spacetime background is given by \cite{NPB, CJPHY, AHEP1, AHEP2, AHEP3, AHEP4, AHEP5}:
\begin{equation}
   \mathcal{L}=\frac{1}{2}\,g_{\mu\nu}\,\dot{x}^{\mu}\,\dot{x}^{\nu},\label{cc1}
\end{equation}
where dot represents differentiation with respect to an affine parameter $\tau$ and $g_{\mu\nu}$ is the metric tensor.

Exploiting the static and spherically symmetric nature of spacetime, we restrict our analysis to geodesic motions in the equatorial plane ($\theta=\pi/2$ and $\dot{\theta}=0$) without loss of generality. The reduced Lagrangian density function (\ref{cc1}) using the spacetime (\ref{bb1}) then takes the following form:
\begin{eqnarray}
\mathcal{L}=\frac{1}{2}\,\Big(-\mathcal{F}(r)\,\dot{t}^2+\mathcal{F}(r)^{-1}\,\dot{r}^2+r^2\,\dot{\phi}^2\Big). \label{cc2}
\end{eqnarray}

The independence of this Lagrangian density function from the coordinates $t$ and $\phi$ yields two conserved quantities corresponding to the energy and the angular momentum given by
\begin{eqnarray}
   \mathrm{E}=\dot{t}\,\mathcal{F}(r),\quad\quad \mathrm{L}=r^2\,\dot{\phi}.\label{cc3}
\end{eqnarray}

From these conservation laws, we derive the radial equation of motion from Eq. (\ref{cc2}) as follows:
\begin{equation}
   \left(\frac{dr}{d\tau}\right)^2+V_\text{eff}(r)=\mathrm{E}^2,\label{cc4}
\end{equation}
where $V_\text{eff}(r)$ is the effective potential governing the radial motion and is given by
\begin{equation}
   V_\text{eff}(r)=\left(-\varepsilon+\frac{\mathrm{L}^2}{r^2}\right)\,\mathcal{F}(r)=\left(-\varepsilon+\frac{\mathrm{L}^2}{r^2}\right)\,\left(1-8\,\pi\,\eta^2\,\xi-\frac{2\,M}{r}+\frac{r^2}{\ell^2_{p}}+\frac{\alpha \beta^2}{3\,r\,(\beta+r)^3}+\frac{\alpha}{(\beta+r)^2} \right),\label{cc5}
\end{equation}
with $\varepsilon=0$ for null geodesics and $\varepsilon=-1$ for timelike geodesics.

From the above expression (\ref{cc5}), it is clear that the effective potential of the system is influenced by several factors. These include the energy scale parameter $\eta$, the deformation parameter $\alpha$, the control parameter $\beta$, and the radius of the curvature $\ell_p$. In the limit when $\alpha=0$, that is, no deformation of the BH, one can recover results for AdS BH with phantom GMs. Noted that this effective potential is more with a phantom GM ($\xi=-1$) compared to an ordinary GM ($\xi=1$). 

\begin{center}
    \underline{\bf Null Geodesics Analysis}
\end{center}

For null geodesics, the effective potential Eq. (\ref{cc5}) simplifies to:
\begin{equation}
   V_\text{eff}(r)=\frac{\mathrm{L}^2}{r^2}\,\mathcal{F}(r)=\frac{\mathrm{L}^2}{r^2}\,\left(1-8\,\pi\,\eta^2\,\xi-\frac{2\,M}{r}+\frac{r^2}{\ell^2_{p}}+\frac{\alpha \beta^2}{3\,r\,(\beta+r)^3}+\frac{\alpha}{(\beta+r)^2} \right).\label{dd1}
\end{equation}

For circular null geodesics, which define the photon sphere, satisfy the conditions $\dot{r}=0$ and $\ddot{r}=0$ at $r=r_c$, leading to the following relations
\begin{equation}
   V_\text{eff}(r_c)=\mathrm{E}^2,\quad\quad V'_\text{eff}(r_c)=0,\label{dd2}
\end{equation}
where prime denotes differentiation with respect to $r$.

The first condition yields the critical impact parameter for photon light given by
\begin{equation}
   \frac{1}{\beta^2_{c}}=\frac{1}{r^2_c}\,\left[1-8\,\pi\,\eta^2\,\xi-\frac{2\,M}{r_c}+\frac{r^2_{c}}{\ell^2_{p}}+\frac{\alpha\,\beta^2}{3\,r_c\,(\beta+r_c)^3}+\frac{\alpha}{(\beta+r_c)^2}\right].\label{dd3}
\end{equation}
From the above expression (\ref{dd3}), it is clear that the critical impact parameter is influenced by several factors. These include the energy scale of the symmetry breaking parameter $\eta$, the deformation parameter $\alpha$, the control parameter $\beta$, and  radius of the curvature $\ell_p$. In the limit when $\alpha=0$, that is, no deformation of the BH, one can recover results for AdS BH with phantom GMs. Noted that this impact parameter is more with a phantom GM ($\xi=-1$) compared to an ordinary GM ($\xi=1$) which is clear from the below analysis. 

In the presence of an ordinary GM ($\xi=1$), this critical impact parameter becomes:
\begin{equation}
   \frac{1}{\beta^2_{c}}=\frac{1}{r^2_c}\,\left[1-8\,\pi\,\eta^2-\frac{2\,M}{r_c}+\frac{r^2_{c}}{\ell^2_{p}}+\frac{\alpha\,\beta^2}{3\,r_c\,(\beta+r_c)^3}+\frac{\alpha}{(\beta+r_c)^2}\right].\label{dd3aa}
\end{equation}
While with a phantom GM ($\xi=-1$), we obtain:
\begin{equation}
   \frac{1}{\beta^2_{c}}=\frac{1}{r^2_c}\,\left[1+8\,\pi\,\eta^2-\frac{2\,M}{r_c}+\frac{r^2_{c}}{\ell^2_{p}}+\frac{\alpha\,\beta^2}{r_c\,(\beta+r_c)^3}+\frac{\alpha}{(\beta+r_c)^2}\right].\label{dd3bb}
\end{equation}
Comparing Eqs.~(\ref{dd3aa}) and (\ref{dd3bb}), we observe that the critical impact parameter is generally larger for a phantom GM than for an ordinary GM. Physically, this means that photon approaching a BH with phantom GM require a larger minimum impact parameter to avoid capture compared to BHs with an ordinary GM. This finding has significant implications for gravitational lensing, suggesting that a phantom GM produce stronger deflection effects for light rays passing at the same distance from the BH. Such differences could potentially be detected through high-precision observations of gravitational lensing events, providing an observational means to distinguish between different types of topological defects in the vicinity of BHs. The second condition in Eq.~(\ref{dd2}) yields the photon sphere radius $r=r_\text{ph}$ which we discussed earlier.

To analyze the forces experienced by photon particle in the gravitational field produced by the selected BH, we compute the radial force derived from the effective potential defined by the following relation:
\begin{equation}
   \mathrm{F}_\text{ph}=-\frac{1}{2}\,\frac{dV_\text{eff}(r)}{dr}.\label{dd5}
\end{equation}
Substituting the effective potential from Eq.~(\ref{dd1}) yields the following expression:
\begin{equation}
   \mathrm{F}_\text{ph}=\frac{\mathrm{L}^2}{r^3}\,\left[1-8\,\pi\,\eta^2\,\xi-\frac{3\,M}{r}+\frac{\alpha}{(\beta+r)^2}+\frac{\alpha\,\beta^2}{2\,r\,(\beta+r)^3}+\frac{\alpha\,r}{(\beta+r)^3}+\frac{\alpha\,\beta^2}{2\,(\beta+r)^4}\right].\label{dd6}
\end{equation}
From the above expression (\ref{dd6}), it is clear that the radial force on photon particle is influenced by several factors. These include the energy scale of the symmetry breaking parameter $\eta$, the deformation parameter $\alpha$, the control parameter $\beta$, and radius of the curvature $\ell_p$. In the limit when $\alpha=0$, that is, no deformation in the BH, one can recover results for AdS BH with phantom global monopole. Noted that this force expression is more with a phantom GM ($\xi=-1$) compared to an ordinary GM ($\xi=1$) which is clear from the below analysis. 

In the presence of an ordinary GM ($\xi=1$) in the selected BH spacetime, the force on photon particle is given by:
\begin{equation}
   \mathrm{F}_\text{ph}=\frac{\mathrm{L}^2}{r^3}\,\left[1-8\,\pi\,\eta^2-\frac{3\,M}{r}+\frac{\alpha}{(\beta+r)^2}+\frac{\alpha\,\beta^2}{2\,r\,(\beta+r)^3}+\frac{\alpha\,r}{(\beta+r)^3}+\frac{\alpha\,\beta}{2\,(\beta+r)^4}\right].\label{dd6aa}
\end{equation}
While with a phantom GM ($\xi=-1$), the radial force becomes:
\begin{equation}
   \mathrm{F}_\text{ph}=\frac{\mathrm{L}^2}{r^3}\,\left[1+8\,\pi\,\eta^2-\frac{3\,M}{r}+\frac{\alpha}{(\beta+r)^2}+\frac{\alpha\,\beta^2}{2\,r\,(\beta+r)^3}+\frac{\alpha\,r}{(\beta+r)^3}+\frac{\alpha\,\beta}{2\,(\beta+r)^4}\right].\label{dd6bb}
\end{equation}
The comparison between Eqs.~(\ref{dd6aa}) and (\ref{dd6bb}) shows a critical distinction: the force term for phantom GMs contains a positive contribution $(1+8\,\pi\,\eta^2)$ compared to the negative contribution $(1-8\,\pi\,\eta^2)$ for an ordinary GM. This difference means that at the same radial distance, photons experience a stronger radial force in spacetime with a phantom GM compared to those with an ordinary GM. The magnitude of this difference scales with $\eta^2$, indicating that the effect becomes more pronounced at higher energy scales of symmetry breaking. This enhanced force explains why a phantom GM produces stronger gravitational lensing effects despite having smaller photon sphere radii-the steeper gradient of the effective potential leads to greater deflection of photon trajectories.

We now analyze the stability of circular null orbits in the equatorial plane. A key physical quantity that determines this stability is the Lyapunov exponent, which, in terms of the effective potential $V_\text{eff}(r)$, is defined as follows:
\begin{equation}
    \lambda^\text{null}_L=\sqrt{-\frac{V''_\text{eff}(r)}{2\,\dot{t}^2}}.\label{stable1}
\end{equation}
Using Eqs. (\ref{cc3}), (\ref{dd1}) and the condition $2\,\mathcal{F}(r)=r\,\mathcal{F}'(r)$, we find the following expression of the Lyapunov exponent {\small
\begin{eqnarray}
    &&\lambda^\text{null}_L=\sqrt{\mathcal{F}(r)\,\left(\frac{\mathcal{F}(r)}{r^2}-\frac{\mathcal{F}''(r)}{2}\right)}\nonumber\\
    &=&\sqrt{1-8\,\pi\,\eta^2\,\xi-\frac{2\,M}{r}+\frac{r^2}{\ell^2_{p}}+\frac{\alpha \beta^2}{3\,r\,(\beta+r)^3}+\frac{\alpha}{(\beta+r)^2}}\,\sqrt{\frac{r^5 - 2\, r^3\, \alpha + 5\, r^4\, \beta + 10\, r^3\, \beta^2 + 
 10\, r^2\, \beta^3 + 
 5\, r\, \beta^4 + \beta^5 -8\,\pi\, (r + \beta)^5\,\eta^2\,\xi}{r^2\, (r + \beta)^5}}\Bigg{|}_{r=r_\text{ph}}.\quad\quad\label{stable2}
\end{eqnarray}
} From the above expression (\ref{stable2}), it is clear that the Lyapunov exponent for circular null geodesics is influenced by several factors. These include the energy scale parameter $\eta$, the deformation parameter $\alpha$, the control parameter $\beta$, the radius of curvature $\ell_p$, and the BH mass $M$. In the limit when $\alpha=0$, that is, without any deformation of the BH, one can recover results for AdS BH with phantom GMs. Moreover, in the limit when $\beta=0$, that is, without the control parameter in the selected BH solution, the Lyapunov exponent from Eq. (\ref{stable2}) becomes
\begin{eqnarray}
\lambda^\text{null}_L=\frac{1}{r}\,\sqrt{\left(1-8\,\pi\,\eta^2\,\xi-\frac{2\,M}{r}+\frac{r^2}{\ell^2_{p}}+\frac{\alpha}{r^2}\right)\,\left(1-8\,\pi\,\eta^2\,\xi-\frac{2\,\alpha}{r^2}\right)}\Bigg{|}_{r=r_\text{ph}}.\label{stable3}
\end{eqnarray}

To further analyze the trajectories of photons, we derive the orbit equation by combining Eqs.~(\ref{cc3}), (\ref{cc4}), and (\ref{dd1}):
\begin{equation}
   \frac{\dot{r}^2}{\dot{\phi}^2}=\left(\frac{dr}{d\phi}\right)^2=r^4\,\left[\frac{1}{\gamma^2}-\frac{1}{r^2}\,\left\{1-8\,\pi\,\eta^2\,\xi-\frac{2\,M}{r}+\frac{r^2}{\ell^2_{p}}+\frac{\gamma_1}{r\,(\beta+r)^3}+\frac{\gamma_2}{(\beta+r)^2}\right\} \right],\label{dd7}
\end{equation}
where $\gamma=\frac{\mathrm{E}}{\mathrm{L}}$ is the impact parameter for photon trajectories.

Introducing the substitution $u=\frac{1}{r}$, we obtain the trajectory equation:
\begin{equation}
   \left(\frac{du}{d\phi}\right)^2+(1-8\,\pi\,\eta^2\,\xi)\,u^2=\frac{1}{\gamma^2}+\frac{1}{\ell^2_{p}}+2\,M\,u^3-{\frac{\alpha\,\beta^2\,u^6}{3\,(1+\beta\,u)^3}-\frac{\alpha\,u^4}{(1+\beta\,u)^2}}.\label{dd8}
\end{equation}
The equation (\ref{dd8}) represents the photon trajectory equation under the gravitational field of a deformed AdS BH with phantom GMs. From the trajectory equation, it is evident that several factors influence the photon paths, including the deformation parameter $\alpha$, the control parameter $\beta$, the impact parameter $\gamma$, the radius of curvature $\ell_p$, and the BH mass $M$. These parameters play a crucial role in shaping the photon trajectories in this gravitational setup.

Thereby, with an ordinary GM ($\xi=1$), this trajectory equation becomes:
\begin{equation}
   \left(\frac{du}{d\phi}\right)^2+(1-8\,\pi\,\eta^2)\,u^2=\frac{1}{\gamma^2}+\frac{1}{\ell^2_{p}}+2\,M\,u^3-{\frac{\alpha\,\beta^2\,u^6}{3\,(1+\beta\,u)^3}-\frac{\alpha\,u^4}{(1+\beta\,u)^2}}.\label{dd9}
\end{equation}
While with a phantom GM ($\xi=-1$), we find:
\begin{equation}
   \left(\frac{du}{d\phi}\right)^2+(1+8\,\pi\,\eta^2)\,u^2=\frac{1}{\gamma^2}+\frac{1}{\ell^2_{p}}+2\,M\,u^3-{\frac{\alpha\,\beta^2\,u^6}{3\,(1+\beta\,u)^3}-\frac{\alpha\,u^4}{(1+\beta\,u)^2}}.\label{dd10}
\end{equation}

The structural difference between Eqs.~(\ref{dd9}) and (\ref{dd10}) lies in the coefficient of the $u^2$ term, which is $(1-8\,\pi\,\eta^2)$ for an ordinary GM and $(1+8\,\pi\,\eta^2)$ for a phantom GM. This term significantly influences the angular motion of photons, with the positive contribution in the phantom GM case enhancing the angular component of photon trajectories. This enhancement leads to greater deflection angles for a phantom GM compared to an ordinary GM, which is consistent with our gravitational lensing analysis.

The trajectory equations also show how the deformation parameter $\alpha$ and the control parameter $\beta$ together influences the photon paths through the complex terms ${\frac{\alpha\,\beta^2\,u^6}{3\,(1+\beta\,u)^3}}$ and ${\frac{\alpha\,u^4}{(1+\beta\,u)^2}}$. These terms introduce radial-dependent corrections to the standard Schwarzschild photon trajectories, with the effect being most significant at smaller radial distances (larger values of $u$). This behavior explains why the impact of deformation parameters on photon orbits becomes more pronounced near the BH, particularly affecting the photon sphere radius and critical impact parameter.

In the limit when $\beta=0$, that is, without control parameter, the trajectory equation from Eq. (\ref{dd8}) becomes
\begin{equation}
   \left(\frac{du}{d\phi}\right)^2+(1-8\,\pi\,\eta^2\,\xi)\,u^2=\frac{1}{\gamma^2}+\frac{1}{\ell^2_{p}}+2\,M\,u^3-\alpha\,u^4.\label{dd11}
\end{equation}
Differentiating w. r. t. $\phi$ and after simplification yields:
\begin{equation}
   \frac{d^2u}{d\phi^2}+(1-8\,\pi\,\eta^2\,\xi)\,u=3\,M\,u^2-2\,\alpha\,u^3\label{dd12}
\end{equation}
which further reduces to the result for the Schwrazschild BH case when $\alpha=0=\xi$.

\begin{center}
    \underline{\bf Timelike Geodesics Analysis}
\end{center}

For timelike geodesics ($\varepsilon=-1$), the effective potential takes the form:
\begin{equation}
   V_\text{eff}(r)=\left(1+\frac{\mathrm{L}^2}{r^2}\right)\,\mathcal{F}(r),\label{ee1}
\end{equation}

Circular timelike geodesics satisfy the conditions $\dot{r}=0$ and $\ddot{r}=0$, yielding:
\begin{equation}
   \mathrm{E}^2=V_\text{eff}(r)=\left(1+\frac{\mathrm{L}^2}{r^2}\right)\,\mathcal{F}(r),\quad\quad  V'_\text{eff}(r)=0.\label{ee2}
\end{equation}

From these conditions, we find the angular momentum of timelike particles in the circular orbits:
\begin{eqnarray}
   \mathrm{L}=r\,\sqrt{\frac{\frac{M}{r}+\frac{r^2}{\ell^2_{p}}-\frac{\alpha\,\beta^2}{6}\,\left(\frac{1}{r\,(\beta+r)^3}+\frac{3}{(\beta+r)^4}\right)-\frac{\alpha\,r}{(\beta+r)^3}}{1-8\,\pi\,\eta^2\,\xi-\frac{3\,M}{r}+\frac{\alpha}{(\beta+r)^2}+\frac{\alpha\,\beta^2}{2\,r\,(\beta+r)^3}+\frac{\alpha\,r}{(\beta+r)^3}+\frac{\alpha\,\beta^2}{2\,(\beta+r)^4}}}.\label{ee4}
\end{eqnarray}
And the particles energy is given by:
\begin{eqnarray}
   \mathrm{E}_{\pm}=\pm\,\frac{1-8\,\pi\,\eta^2\,\xi-\frac{2\,M}{r}+\frac{r^2}{\ell^2_{p}}+\frac{\alpha\,\beta^2}{3\,r\,(\beta+r)^3}+\frac{\alpha}{(\beta+r)^2}}{\sqrt{1-8\,\pi\,\eta^2\,\xi-\frac{3\,M}{r}+\frac{\alpha}{(\beta+r)^2}+\frac{\alpha\,\beta^2}{2\,r\,(\beta+r)^3}+\frac{\alpha\,r}{(\beta+r)^3}+\frac{\alpha\,\beta^2}{2\,(\beta+r)^4}}}.\label{ee5}
\end{eqnarray}
From the expressions (\ref{ee4})--(\ref{ee5}), it is evident that the physical quantities ($\mathrm{L},\, \mathrm{E}_{\pm}$) associated with timelike particles orbiting on circular geodesics in the equatorial plane are influenced by several factors. These factors include the energy scale of the symmetry-breaking parameter $\eta$, the deformation parameter $\alpha$, the control parameter $\beta$, the radius of curvature $\ell_p$, and the BH mass $M$. It is important to note that the presence of a phantom global monopole ($\xi=-1$) and an ordinary global monopole ($\xi=1$) individually affect these physical quantities for timelike particles.

In the presence of an ordinary GM ($\xi=1$), these physical quantity becomes:
\begin{eqnarray}
   &&\mathrm{L}^\text{ordinary GM}=r\,\sqrt{\frac{\frac{M}{r}+\frac{r^2}{\ell^2_{p}}-\frac{\alpha\,\beta^2}{6}\,\left(\frac{1}{r\,(\beta+r)^3}+\frac{3}{(\beta+r)^4}\right)-\frac{\alpha\,r}{(\beta+r)^3}}{1-\eta^2-\frac{3\,M}{r}+\frac{\alpha}{(\beta+r)^2}+\frac{\alpha\,\beta^2}{2\,r\,(\beta+r)^3}+\frac{\alpha\,r}{(\beta+r)^3}+\frac{\alpha\,\beta^2}{2\,(\beta+r)^4}}},\nonumber\\
   &&\mathrm{E}^\text{ordinary GM}_{\pm}=\pm\,\frac{1-8\,\pi\,\eta^2-\frac{2\,M}{r}+\frac{r^2}{\ell^2_{p}}+\frac{\alpha\,\beta^2}{3\,r\,(\beta+r)^3}+\frac{\alpha}{(\beta+r)^2}}{\sqrt{1-8\,\pi\,\eta^2-\frac{3\,M}{r}+\frac{\alpha}{(\beta+r)^2}+\frac{\alpha\,\beta^2}{2\,r\,(\beta+r)^3}+\frac{\alpha\,r}{(\beta+r)^3}+\frac{\alpha\,\beta^2}{2\,(\beta+r)^4}}}.\label{ee5aa}
\end{eqnarray}
While with a phantom GM ($\xi=-1$), we have:
\begin{eqnarray}
   &&\mathrm{L}^\text{phantom GM}=r\,\sqrt{\frac{\frac{M}{r}+\frac{r^2}{\ell^2_{p}}-\frac{\alpha\,\beta^2}{6}\,\left(\frac{1}{r\,(\beta+r)^3}+\frac{3}{(\beta+r)^4}\right)-\frac{\alpha\,r}{(\beta+r)^3}}{1+\eta^2-\frac{3\,M}{r}+\frac{\alpha}{(\beta+r)^2}+\frac{\alpha\,\beta^2}{2\,r\,(\beta+r)^3}+\frac{\alpha\,r}{(\beta+r)^3}+\frac{\alpha\,\beta^2}{2\,(\beta+r)^4}}},\nonumber\\
   &&\mathrm{E}^\text{ordinary GM}_{\pm}=\pm\,\frac{1+8\,\pi\,\eta^2-\frac{2\,M}{r}+\frac{r^2}{\ell^2_{p}}+\frac{\alpha\,\beta^2}{3\,r\,(\beta+r)^3}+\frac{\alpha}{(\beta+r)^2}}{\sqrt{1+8\,\pi\,\eta^2-\frac{3\,M}{r}+\frac{\alpha}{(\beta+r)^2}+\frac{\alpha\,\beta^2}{2\,r\,(\beta+r)^3}+\frac{\alpha\,r}{(\beta+r)^3}+\frac{\alpha\,\beta^2}{2\,(\beta+r)^4}}}.\label{ee5bb}
\end{eqnarray}
From the expressions (\ref{ee4})--(\ref{ee5}), it is evident that $\mathrm{L}^\text{ordinary GM}>\mathrm{L}^\text{phantom GM}$ indicating higher angular momentum of timelike particle orbiting on circular geodesics in the equatorial plane in the presence of an ordinary global monopole compared with a phantom global monopole. In contrast, the particles' energy is more in the presence of a phantom global monopole compared to an ordinary global monopole, that is, $\mathrm{E}^\text{phantom GM}_{\pm}> \mathrm{E}^\text{ordinary GM}_{\pm}$.

We discuss a special case corresponds to $\beta=0$, that is, without the control parameter in the BH solution. The above physical quantities for an ordinary and a phantom GMs become:
\begin{eqnarray}
&&\mathrm{L}^\text{ordinary GM}=r\,\sqrt{\frac{\frac{M}{r}+\frac{r^2}{\ell^2_{p}}-\frac{\alpha}{r^2}}{1-8\,\pi\,\eta^2-\frac{3\,M}{r}+\frac{2\,\alpha}{r^2}}},\quad\quad\quad\,\, \mathrm{L}^\text{phantom GM}=r\,\sqrt{\frac{\frac{M}{r}+\frac{r^2}{\ell^2_{p}}-\frac{\alpha}{r^2}}{1+8\,\pi\,\eta^2-\frac{3\,M}{r}+\frac{2\,\alpha}{r^2}}}\nonumber\\
&&\mathrm{E}^\text{ordinary GM}_{\pm}=\pm\,\frac{1-8\,\pi\,\eta^2-\frac{2\,M}{r}+\frac{r^2}{\ell^2_{p}}+\frac{\alpha}{r^2}}{\sqrt{1-8\,\pi\,\eta^2-\frac{3\,M}{r}+\frac{2\,\alpha}{r^2}}},\quad\quad \mathrm{E}^\text{phantom GM}_{\pm}=\pm\,\frac{1+8\,\pi\,\eta^2-\frac{2\,M}{r}+\frac{r^2}{\ell^2_{p}}+\frac{\alpha}{r^2}}{\sqrt{1+8\,\pi\,\eta^2-\frac{3\,M}{r}+\frac{2\,\alpha}{r^2}}}.\label{ee5cc}
\end{eqnarray}
The radius of photon sphere in that case becomes 
\begin{equation}
r_\text{ph}=\frac{3\,M}{2\,(1-8\,\pi\,\eta^2\,\xi)}\,\left[1+\sqrt{1-\frac{8\,\alpha}{9\,M^2}\,(1-8\,\pi\,\eta^2\,\xi)}\right] \simeq     \frac{3\,M}{(1-8\,\pi\,\eta^2\,\xi)}-\frac{2\,\alpha}{3\,M}.\label{ee5dd}
\end{equation}

Now, we aim to find speed with which timelike particle orbits the BH at a very large distance in comparison with the horizon of the BH. This is in analogy with a distant star in a galaxy moving in a circle around the BH of the galaxy. In the zeroth approximation, one write
\begin{equation}
    \mathcal{F}(r)=1+2\,\Phi(r)\label{pp1}
\end{equation}
in which $\Phi(r)$ is the Newtonian gravitational potential for the timelike particle of unit mass. Therefore, explicitly, we get
\begin{equation}
    \Phi(r)=\frac{1}{2}\,\left[-8\,\pi\,\eta^2\,\xi-\frac{2\,M}{r}+\frac{r^2}{\ell^2_{p}}+\frac{\alpha\, \beta^2}{3\,r\,(\beta+r)^3}+\frac{\alpha}{(\beta+r)^2}\right].\label{pp2}
\end{equation}
Now, the effective gravitational force is simply given by $\mathrm{F}_c=-\frac{\partial \Phi(r)}{\partial r}$ which is toward the center of the BH. We find this effective force given by
\begin{eqnarray}
    \mathrm{F}_c=\left[-\frac{M}{r^2}-\frac{r}{\ell^2_{p}}+\frac{\alpha\,\beta^2}{6\,r^2\,(\beta+r)^3}+\frac{\alpha\,\beta^2}{2\,r\,(\beta+r)^4}+\frac{\alpha}{(\beta+r)^3}\right].\label{pp3}
\end{eqnarray}
This central force can be equated with centripetal acceleration, {\it i.e.}, $|\mathrm{F}_c|=\frac{v^2}{r}$ in which $v$ is the speed of the test particle in the orbit. This equation results an expression for the circular speed $v$ given by
\begin{eqnarray}
    v=\sqrt{r\left|-\frac{M}{r^2}-\frac{r}{\ell^2_{p}}+\frac{\alpha\,\beta^2}{6\,r^2\,(\beta+r)^3}+\frac{\alpha\,\beta^2}{2\,r\,(\beta+r)^4}+\frac{\alpha}{(\beta+r)^3}\right|}.\label{pp4}
\end{eqnarray}

The expression (\ref{pp4}) shows that the circular speed of timelike particles in orbit is affected by several parameters. These include the deformation parameter $\alpha$, the control parameter $\beta$, and the radius of curvature $\ell_p$. 

In the limit where $\beta=0$, that is without the control parameter in the BH solution, then the selected BH solution reduces an AdS BH metric with quantum-like potential, reported in \cite{AFA}. In that case, we find the speed of timelike particles on the circular orbits becomes:  
\begin{eqnarray}
    v=\sqrt{\left|\frac{\alpha}{r^2}-\frac{M}{r}-\frac{r^2}{\ell^2_{p}}\right|}.\label{pp5}
\end{eqnarray}

A critical feature of timelike geodesics is the innermost stable circular orbit (ISCO) \cite{isz91,isz92,isz93,isz94,isz95}, which marks the transition between stable and unstable circular motion. The ISCO radius is determined by solving:
\begin{equation}
V_\text{eff}=0,\quad\quad V_\text{eff}^{\prime }=0,\quad\quad V_\text{eff}^{\prime \prime }\geq 0,  \label{veff44}
\end{equation}
where $V_\text{eff}$ is given by Eq.~(\ref{ee1}).

These conditions lead to the ISCO equation:
\begin{equation}
\frac{4\,\mathcal{F}\,\mathcal{F}'}{r}-\frac{\mathcal{F}\,\mathcal{F}'}{r}+\mathcal{F}\,\mathcal{F}''-2\,\mathcal{F}'^2=0.
\label{isco} 
\end{equation}

This equation cannot be solved analytically, necessitating numerical methods. Our numerical results for the ISCO radius are presented in Tables \ref{taba14} and \ref{taba15}. Table \ref{taba14} shows that for ordinary GMs ($\xi=1$), the ISCO radius decreases with increasing values of $\eta$. For example, at fixed values $\alpha=0.2$ and $\beta=0.2$, the ISCO radius decreases from $r_{ISCO}=3.46395$ at $\eta=0.2$ to $r_{ISCO}=3.09294$ at $\eta=0.4$, and further contracts to $r_{ISCO}=2.62081$ at $\eta=0.6$, representing a reduction of approximately $24\%$ across this parameter range.

\begin{center}
\begin{tabular}{|c|c|c|c|c|c|c|c|c|c|c|}
 \hline 
 \multicolumn{10}{|c|}{ $r_{ISCO}$ for ordinary GMs}
\\ \hline 
$\zeta =1$ & \multicolumn{3}{|c|}{ $\eta =0.2$ }& \multicolumn{3}{|c|}{ $\eta =0.4$ } & \multicolumn{3}{|c|}{ $\eta =0.6$ }  \\ \hline
$\alpha $ & $\beta =0.2$ & $0.4$ & $0.6$ & $\beta =0.2$ & $0.4$ & $0.6$ & $%
\beta =0.2$ & $0.4$ & $0.6$ \\ \hline
$0.2$ & $3.46395$ & $3.4838$ & $3.49936$ & $3.09294$ & $3.11438$ & $3.13077$
& $2.62081$ & $2.64463$ & $2.66211$ \\ 
$0.4$ & $3.31112$ & $3.35516$ & $3.38873$ & $2.94125$ & $2.98921$ & $3.02467$
& $2.47116$ & $2.52508$ & $2.56302$ \\ 
$0.6$ & $3.14426$ & $3.21865$ & $3.27335$ & $2.77407$ & $2.85597$ & $2.91394$
& $2.30371$ & $2.39733$ & $2.45961$\\ 
 \hline
\end{tabular}
\captionof{table}{Numerical results for the ISCO in case of ordinary GMs with various BH parameters $\alpha$ and $\beta$. Here $M=1$ and $\Lambda=-0.002$.} \label{taba14}
\end{center}
\begin{center}
\begin{tabular}{|c|c|c|c|c|c|c|c|c|c|c|}
 \hline 
 \multicolumn{10}{|c|}{ $r_{ISCO}$ for phantom GMs}
\\ \hline 
$\zeta =-1$ & \multicolumn{3}{|c|}{ $\eta =0.2$ }& \multicolumn{3}{|c|}{ $\eta =0.4$ } & \multicolumn{3}{|c|}{ $\eta =0.6$ }  \\ \hline
$\alpha $ & $\beta =0.2$ & $0.4$ & $0.6$ & $\beta =0.2$ & $0.4$ & $0.6$ & $%
\beta =0.2$ & $0.4$ & $0.6$ \\ \hline
$0.2$ & $3.76308$ & $3.7818$ & $3.79673$ & $4.31908$ & $4.33599$ & $4.34984$
& $5.71072$ & $5.7243$ & $5.73594$ \\ 
$0.4$ & $3.60953$ & $3.65081$ & $3.68294$ & $4.16448$ & $4.20143$ & $4.23109$
& $5.55454$ & $5.58369$ & $5.60837$ \\ 
$0.6$ & $3.44295$ & $3.51215$ & $3.56435$ & $3.99841$ & $4.0596$ & $4.10754$
& $5.38955$ & $5.4368$ & $5.47622$\\ 
 \hline
\end{tabular}
\captionof{table}{Numerical results for the ISCO in case of phantom GMs with various BH parameters $\alpha$ and $\beta$. Here $M=1$ and $\Lambda=-0.002$.} \label{taba15}
\end{center}

In striking contrast, Table \ref{taba15} demonstrates that for phantom GMs ($\xi=-1$), the ISCO radius increases dramatically with increasing values of $\eta$. Following the same parameter values $\alpha=0.2$ and $\beta=0.2$, we observe that $r_{ISCO}$ increases from $3.76308$ at $\eta=0.2$ to $4.31908$ at $\eta=0.4$, and expands further to $5.71072$ at $\eta=0.6$. This represents a substantial growth of approximately $52\%$ across the same parameter range. This behavior is particularly noteworthy because it is exactly opposite to what we observed for photon spheres, where phantom GMs led to smaller radii while ordinary GMs resulted in larger radii.

\begin{figure}
    \centering
    \includegraphics[width=0.6\linewidth]{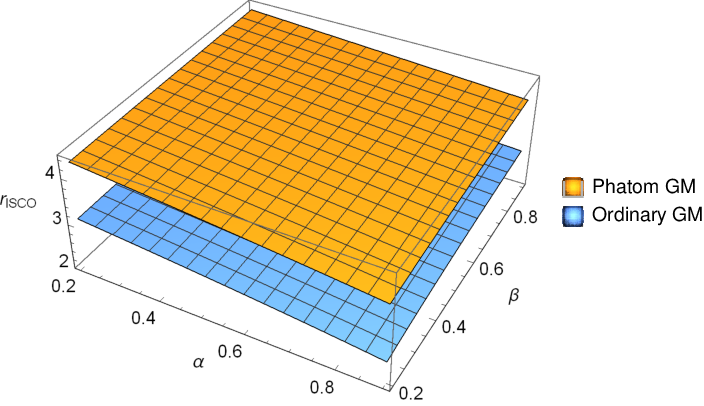}
    \caption{The profile of the ISCO for various values of BH parameters $\alpha$  and $\beta$ for fixed $\eta=0.4$ showing that the $r_{ISCO}$ has higher value for phantom GMs than ordinary GMs. Here $M=1$.}
    \label{figa19}
\end{figure}
\begin{figure}[ht!]
    \centering
    \includegraphics[width=0.45\linewidth]{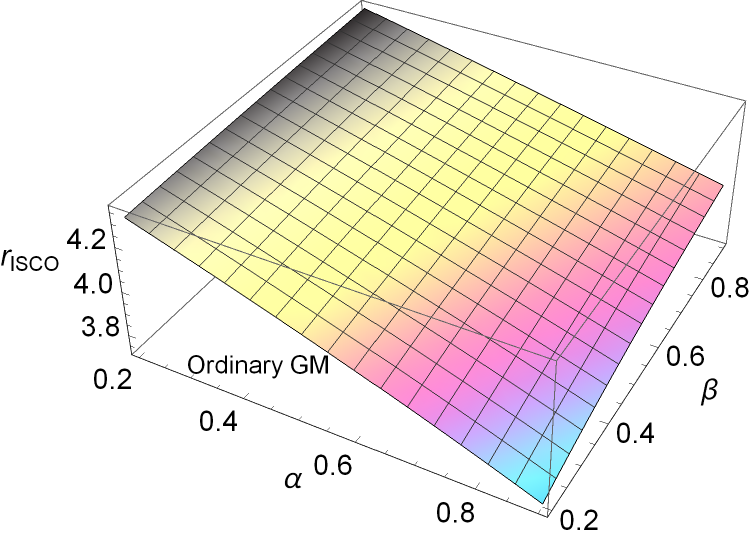}\quad\quad
    \includegraphics[width=0.45\linewidth]{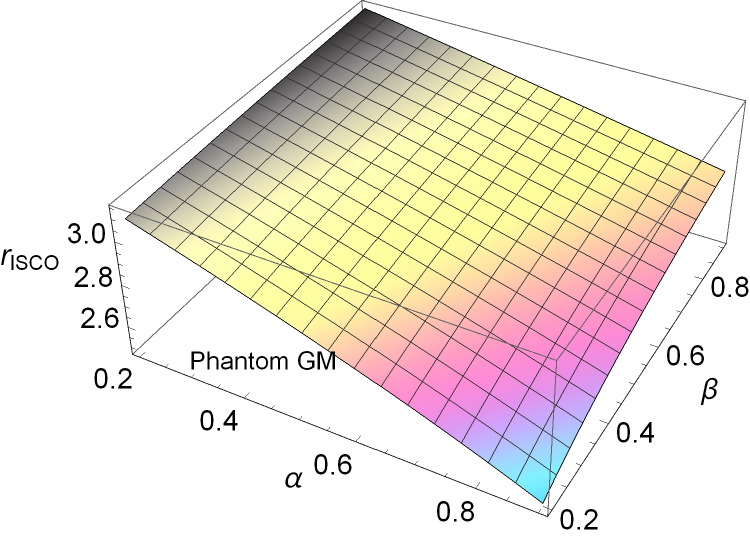}
    \caption{The profile of the  ISCO for various values of BH parameters $\alpha$  and $\beta$ for fixed $\eta=0.4$ showing that the $r_{ISCO}$ decreases with $\alpha$ but increases with $\beta$. Here, $M=1$.}
    \label{figa1b}
\end{figure} 
Figure \ref{figa19} provides a clear visualization of this inverse relationship between ISCO radii for ordinary versus phantom GMs. The upper surface, corresponding to phantom GMs, consistently exhibits larger ISCO radii compared to the lower surface representing ordinary GMs. This topological separation persists throughout the parameter space, confirming the fundamental difference in how these two classes of topological defects influence timelike geodesics. Figure \ref{figa1b} displays the profile of the $r_{ISCO}$ for various values of BH parameters $\alpha$ and $\beta$ with fixed $\eta=0.6$. The $r_{ISCO}$ drops with $\alpha$ but grows with $\beta$.

This contrasting behavior between null and timelike geodesics has profound physical implications. For massive particles, phantom GMs push stable orbits outward, creating a larger region of stability around the BH. Simultaneously, for photons, phantom GMs pull the photon sphere inward, modifying the BH's optical properties and shadow size. This dual effect could potentially provide a distinctive observational signature for detecting phantom GMs in astrophysical environments, as it would manifest in both the matter distribution around the BH (affected by the ISCO) and the electromagnetic observations (gravitational deflection patterns (affected by the photon sphere). This dual signature would be difficult to explain through alternative physical mechanisms, making it a potentially unique indicator of phantom topological defects.

The stability of circular timelike orbits can be further analyzed using the Lyapunov exponent, defined as:
\begin{equation}
   \lambda^\text{timelike}_{L}=\sqrt{(\mathcal{F}'(r))^2-\frac{\mathcal{F}(r)\,\mathcal{F}''(r)}{2}-\frac{3\,\mathcal{F}(r)\,\mathcal{F}'(r)}{2\,r}}.\label{ee8}
\end{equation}

Using the metric function (\ref{bb2}) and setting $\frac{1}{\ell^2_{p}}=-\frac{\Lambda}{3}$, we obtain an explicit, albeit complex, expression for the Lyapunov exponent $\lambda_L$:
\begin{eqnarray}
   \lambda^\text{timelike}_{L}&=&\frac{1}{3\, r^2\, (r + \beta)^4}\,\Bigg[\frac{3}{2}\, (r + \beta)\, \Big\{\alpha\,\beta^2 + 3\, r\, \alpha\,(r + \beta)-6\,M\, (r + \beta)^3 + 3\, r\, (r + \beta)^3-3\, r\, (r + \beta)^3\, 8\,\pi\,\eta^2\,\xi -r^3\, (r + \beta)^3\,\Lambda\Big\}\times\nonumber\\
   &&\Big\{-6\, M\, (r + \beta)^4 + \alpha\,(6\, r^3 + 6\, r^2\, \beta + 
    4\, r\, \beta^2 + \beta^3) + 
 2\, r^3\, (r + \beta)^4\, \Lambda \Big\}\nonumber\\
 &+&\Big\{-6\, M\, (r + \beta)^4 + \alpha\,(6\, r^3 + 6\, r^2\, \beta + 
     4\, r\, \beta^2 + \beta^3) + 
  2\, r^3\, (r + \beta)^4\, \Lambda\Big\}^2\nonumber\\
  &-&\Big\{\alpha\,\beta^2 + 3\, r\, \alpha\,(r + \beta) - 
  6\, M\, (r + \beta)^3 + 3\, r\, (r + \beta)^3 - 
  3\, r\, (r + \beta)^3\, 8\,\pi\,\eta^2\,\xi - 
  r^3\, (r + \beta)^3\, \Lambda\Big\}\times\nonumber\\
  &&\Big\{-6\, M\, (r + \beta)^5 + \alpha\,\Big(9\, r^4 + 9\, r^3 \beta + 
    10\, r^2\, \beta^2 + 5\, r\, \beta^3 + \beta^4\Big) - 
 r^3\, (r + \beta)^5\, \Lambda\Big\}\Bigg]^{1/2}.\label{ee9}
\end{eqnarray}

This expression, while mathematically intricate, encodes crucial information about orbital stability. Positive values of $\lambda_L$ indicate unstable orbits, while negative values correspond to stable configurations. Our numerical analysis reveals that phantom GMs generally increase the Lyapunov exponent compared to ordinary GMs at the same orbital radius, indicating a destabilizing effect on circular orbits. This destabilization is consistent with the repulsive gravitational character of phantom GMs, which weakens the binding between massive particles and the central BH.

In the limit of vanishing deformation ($\alpha=0$), the Lyapunov exponent \cite{isz96,isz97,isz98} reduces to:
\begin{eqnarray}
   \lambda^\text{timelike}_{L}=\sqrt{\frac{6\, M^2}{r^4}+\frac{4}{3}\,(1-8\,\pi\,\eta^2\,\xi)\, \Lambda+\frac{M\, (-1 + 8\,\pi\,\eta^2\,\xi - 5\, r^2\, \Lambda)}{r^3}}.\label{ee10}
\end{eqnarray}

This simplified expression clearly illustrates how the GM parameter $\eta$ and its character (ordinary versus phantom) influence orbital stability in the absence of metric deformations. For ordinary GMs ($\xi=1$), the term $(1-8\pi\eta^2)$ reduces the coefficient of $\Lambda$, decreasing the destabilizing effect of the cosmological constant. Conversely, for phantom GMs ($\xi=-1$), the term $(1+8\pi\eta^2)$ enhances the coefficient of $\Lambda$, amplifying the destabilizing effect. 

Additionally, the term $8\pi\eta^2\xi$ in the numerator of the third term under the square root has opposite signs for ordinary versus phantom GMs. For ordinary GMs, this term reduces the magnitude of the negative contribution, promoting stability, while for phantom GMs, it enhances the negative contribution, further destabilizing the orbit. These mathematical findings align with our physical understanding that phantom GMs introduce repulsive gravitational effects that counteract the attractive forces maintaining stable orbits.

The combined analysis of photon spheres, ISCOs, and Lyapunov exponents reveals a comprehensive picture of how ordinary and phantom GMs differentially affect the geodesic structure of deformed AdS-Schwarzschild BHs. Ordinary GMs enhance the effective gravitational attraction, expanding photon spheres while contracting ISCOs—consistent with strengthening the gravitational field. Phantom GMs produce the opposite effect, contracting photon spheres while expanding ISCOs—consistent with a repulsive contribution that weakens the effective gravitational field. These distinct signatures could potentially be detected through astronomical observations, providing a means to identify and characterize topological defects in the vicinity of astrophysical BHs.

\section{Deflection Angle of Deformed AdS-Schwarzschild BHs with Phantom GM} \label{sec:3}

In this section, we investigate the gravitational lensing properties of deformed AdS-Schwarzschild BHs with phantom   GMs through a rigorous analysis of the weak deflection angle. Gravitational lensing represents one of the most powerful experimental tests of GR, providing direct access to the spacetime curvature around compact objects \cite{isz51}. The presence of both deformation parameters and topological defects introduces distinctive modifications to the lensing behavior, potentially yielding observable signatures that could discriminate between different theoretical models.

We employ the GBTh to derive the deflection angle, a technique that offers significant computational advantages over traditional methods by relating the deflection to the integral of the Gaussian curvature over a suitable domain \cite{isz52}. This approach is particularly well-suited for analyzing modified spacetimes, as it provides a clear geometrical interpretation of how metric deformations influence light propagation.

\subsection{Deflection Angle Calculation Using the Gauss-Bonnet Theorem}\label{sec:3.1}

We begin our analysis within the weak-field approximation framework, considering a spherically symmetric and static BH solution in a non-plasma medium. The GBTh establishes a fundamental connection between the intrinsic curvature of spacetime and the global topological characteristics of the region $\mho_{\cal R}$ bounded by $\partial \mho_{\cal R}$. This relationship is expressed through the following mathematical formulation \cite{isz53}:
\begin{equation}\label{new_eq1}
\iint_{\Sigma_{K}} \mathcal{R} d\mathcal{A}_{2D} + \oint_{\partial \Sigma_{\mathcal{R}}} g_C dt + \sum_{n} \vartheta_n = 2 \pi  \chi \left( \Sigma_{\mathcal{R}} \right),
\end{equation}
where $\Sigma_{\mathcal{R}} \subset \mathcal{A}_{2D}$ is a compact domain within a two-dimensional differentiable surface $\mathcal{A}_{difs}$, bounded by a smooth and oriented contour $\partial \Sigma_{\mathcal{R}}$. Here, $g_C$ denotes the geodesic curvature of $\partial \Sigma_{\mathcal{R}}$, defined as $g_C = \tilde{g} \left( \nabla_{\dot{\lambda}} \dot{\lambda}, \ddot{\lambda} \right)$, where $\tilde{g} ( \dot{\lambda}, \dot{\lambda} ) = 1$ and $\ddot{\lambda}$ is the unit acceleration vector. Additionally, $\vartheta_n$ represents the exterior angle at the $n^{\text{th}}$ vertex of the boundary, and $\chi \left( \Sigma_{\mathcal{R}} \right)$ corresponds to the Euler characteristic number \cite{isz54}. The quantity $K$ refers to the Gaussian optical curvature \cite{isz55}.

To determine the Gaussian optical curvature $K$, we examine null geodesics influenced by the BH \cite{isz56}. Since light propagates along null geodesics (i.e., $ds^2=0$), these paths naturally define an optical metric describing the effective Riemannian geometry experienced by light rays. By applying the null condition and restricting motion to the equatorial plane ($\vartheta = \pi/2$), where the optical metric provides a natural framework for rotational symmetry, we obtain the following optical metric in the transformed coordinate system:

\begin{equation}\label{is4}
dt^2=\tilde{g}_{kl} d\mathrm{x}^k d\mathrm{x}^l= dr_{*}^2+\mathcal{F}^2(r_{*})d\phi^2,
\end{equation}
where 
\begin{equation} \label{is4n}
F(r_{*}(r))=\frac{r}{\sqrt{\mathcal{F}(r)}},
\end{equation}
and $r_{*}$ is the tortoise coordinate \cite{isz57} given by
\begin{equation}\label{is5}
r_{*}=\int{\frac{dr}{F(r)}}.
\end{equation}

The metric's non-zero Christoffel symbols \cite{isz58} are calculated as follows:
\begin{eqnarray} \label{is7.1}
\Gamma_{\phi \phi}^{r_{*}}&=&-F(r_{*})\frac{\mathrm{d}F(r_{\star})}{\mathrm{d}{r_{\star}}},\\
\Gamma_{r_{*}\phi}^{\phi}&=&\frac{1}{F(r_{*})}\frac{\mathrm{d}F(r_{\star})}{\mathrm{d}{r_{\star}}}, \label{is7.2}
\end{eqnarray}
The determinant is given by $\det\tilde{g}_{kl}=F^{2}(r^{\star})$. Thus, the Gaussian optical curvature $K$ is calculated as:
\begin{equation}
K=-\frac{R_{r_{*}\phi r_{*}\phi}}{\det[\tilde{g}_{r\phi}]}=-\frac{1}{F(r_{\star})}\frac{\mathrm{d}^{2}F(r_{\star})}{\mathrm{d}{r_{\star}}^{2}}.
\end{equation}

The expression for optical curvature $K$ can alternatively be reformulated using the variable $r$ \cite{isz59}:
\begin{eqnarray}
K & = &-\frac{1}{F(r^{\star})}\left[\frac{\mathrm{d}r}{\mathrm{d}r^{\star}}\frac{\mathrm{d}}{\mathrm{d}r}\left(\frac{\mathrm{d}r}{\mathrm{d}r^{\star}}\right)\frac{\mathrm{d}F(r)}{\mathrm{d}r}+\left(\frac{\mathrm{d}r}{\mathrm{d}r^{\star}}\right)^{2}\frac{\mathrm{d}^{2}F(r)}{\mathrm{d}r^{2}}\right].\label{is8}
\end{eqnarray}

After substituting the metric function from Eq. (\ref{bb2}) and performing a series expansion for weak-field approximation, we obtain the Gaussian optical curvature:
\begin{equation} \label{is42}
K \approx \frac{3 M^2}{r^4}+\left(-\frac{6}{\ell^2_{p} r}-\frac{2 y}{r^3}\right) M+\frac{y}{\ell^2_{p}}+\frac{6 \alpha}{\ell^2_{p} r^2}-\frac{20 \alpha \beta}{\ell^2_{p} r^3}+\frac{3 \alpha y}{r^4}+\frac{50 \alpha \beta^2}{\ell^2_{p} r^4}+\mathrm{O}\left(\frac{1}{r^5}\right),
\end{equation}
where 
\begin{equation}
   y=1-8\,\pi\,\eta^2\,\xi.
\end{equation}

This expression reveals how the Gaussian optical curvature is influenced by several parameters, including the AdS radius $\ell_p$, deformation parameter $\alpha$, control parameter $\beta$, GM parameter $\xi$, and BH mass $M$. The parameter $y$ encapsulates the contribution from the GM, with $y=1-8\pi\eta^2$ for ordinary GMs ($\xi=1$) and $y=1+8\pi\eta^2$ for phantom GMs ($\xi=-1$). This difference significantly affects the optical curvature and consequently the gravitational lensing behavior.

To calculate the deflection angle, we consider the limit $\cal{R} \rightarrow \infty$ in Eq. (\ref{new_eq1}), where the angle displacement approaches $\pi/2$, ensuring that $\theta_{observer}+\theta_{source}=\pi$, and the Euler characteristic number equals one \cite{isz60}. This yields:

\begin{equation}\label{is43a}
\iint_{\Sigma_{K}} K d\mathcal{A}_{2D} + \oint_{\partial \Sigma_{\mathcal{R}}} g_C dt +\vartheta_n = 2 \pi  \chi \left( \Sigma_{\mathcal{R}} \right),
\end{equation}
where $\alpha_{z}$ denotes the aggregate angle of the jumps, established as $\pi$. 

For high values of $\mathcal{R}$, treating $Q_{\cal R}:=r(\phi)={\cal R}$ as a constant, we derive $\left(\dot{Q}_{R}^{\phi}\right)^{2}=F^{-2}(r_{*})$. The geodesic curvature is subsequently determined as:
\begin{equation}\label{is45}
\left(\nabla_{\dot{Q}_{\cal R}^{r}} \dot{Q}_{{\cal R}}^{r}\right)^{r} \rightarrow \frac{1}{{\cal R}},
\end{equation}
implying that $g_{C}\left(Q_{\mathcal{R}}\right) \to \frac{1}{\mathcal{R}}$. Using the optical metric from Eq. (\ref{is4}) and $d t={\cal R} d \phi$, we obtain:

\begin{equation}\label{is46}
g_{C}(Q_{\cal R})dt=\lim_{{\cal R}\to\infty}[g_{C}(Q_{\cal R})dt]
        =\lim_{{\cal R}\to\infty}\left[\sqrt{\frac{ \tilde{g}^{\phi\phi}}{4\tilde{g}_{r_{*}r_{*}} }}\left(\frac{\partial\tilde{g}_{\phi\phi}}{\partial r_{*}}\right)\right]d\phi
        =d\phi.
\end{equation}

Considering all previous findings, the GBTh yields:
\begin{equation}\label{is47}
\iint_{\Sigma_{K}} K d\mathcal{A}_{2D} + \oint_{\partial \Sigma_{\mathcal{R}}} g_C dt\stackrel{{\cal R}\rightarrow \infty} {=}  \iint_{\mathcal {A}_{difs:\infty}} K d \mathcal{A}_{difs}+\int_{0}^{\pi+\tilde{\Delta}} d \phi.
\end{equation}

Within the framework of the weak deflection limit, the path of a light ray can be approximated as a linear trajectory, expressed by $r(t)\equiv{\mathfrak{D}}=\frac{b}{\sin\varphi}$, with $b$ serving as the impact parameter \cite{isz61}. The deflection angle $\tilde{\delta}$ is then determined using:

\begin{equation}\label{is48}
\tilde{\Delta}=-\int_{0}^{\pi} \int_{\mathfrak{D}}^{\infty} K d\mathcal{A}_{difs}=-\int_{0}^{\pi} \int_{\mathfrak{D}}^{\infty}\frac{K \sqrt{\operatorname{det} \tilde{g}}}{\mathcal{F}(r)} dr d \phi =-\int_{0}^{\pi} \int_{\mathfrak{D}}^{\infty}  \frac{rK}{\mathcal{F}(r)^\frac{3}{2}} d r d \phi.
\end{equation}

Substituting the Gaussian optical curvature from Eq. (\ref{is42}) into Eq. (\ref{is48}) and integrating, we obtain the deflection angle:
\begin{equation} \label{is50}
\tilde{\Delta}\approx\left(\frac{2 y^2}{3 b^3} - \frac{15 M y \pi}{16 b^4}\right) \ell_p^3 + \left(-\frac{2 y}{b} + \frac{15 \alpha \beta \pi}{8 b^4} - \frac{8 \alpha}{3 b^3} + \frac{3 \pi M}{2 b^2}\right) \ell_p.
\end{equation}

This analytical expression explicitly demonstrates how the deflection angle is influenced by both the BH parameters ($M$, $\ell_p$) and the deformation/GM parameters ($\alpha$, $\beta$, $\eta$, $\xi$). The parameter $y$ appears prominently, highlighting the significant difference in lensing behavior between ordinary and phantom GMs. For ordinary GMs ($\xi=1$), $y=1-8\pi\eta^2$, which reduces the coefficient of the $\ell_p^3$ term, while for phantom GMs ($\xi=-1$), $y=1+8\pi\eta^2$, which enhances this coefficient. This fundamental difference leads to stronger deflection angles for phantom GMs compared to ordinary GMs, a distinctive signature that could potentially be observed in gravitational lensing events.

\subsection{Numerical Analysis and Observational Implications}\label{sec:3.2}

\begin{figure*}
    \centering
{\includegraphics[scale=0.55]{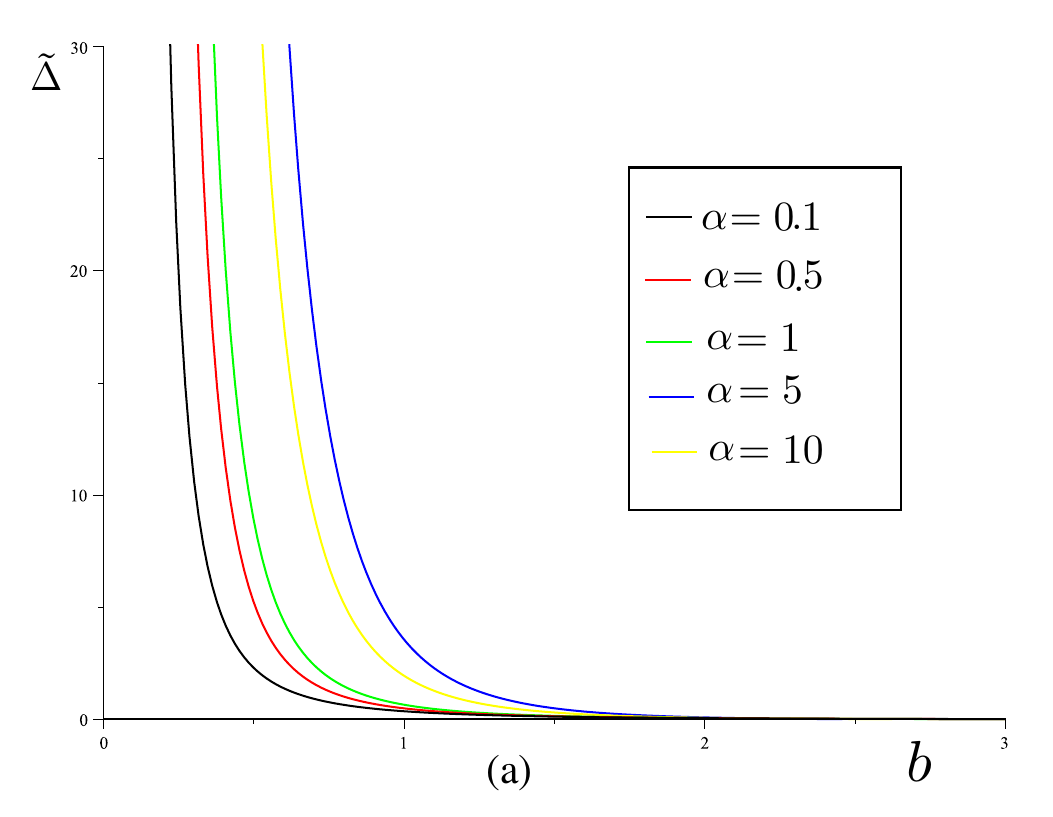} }\qquad
    {{\includegraphics[scale=0.55]{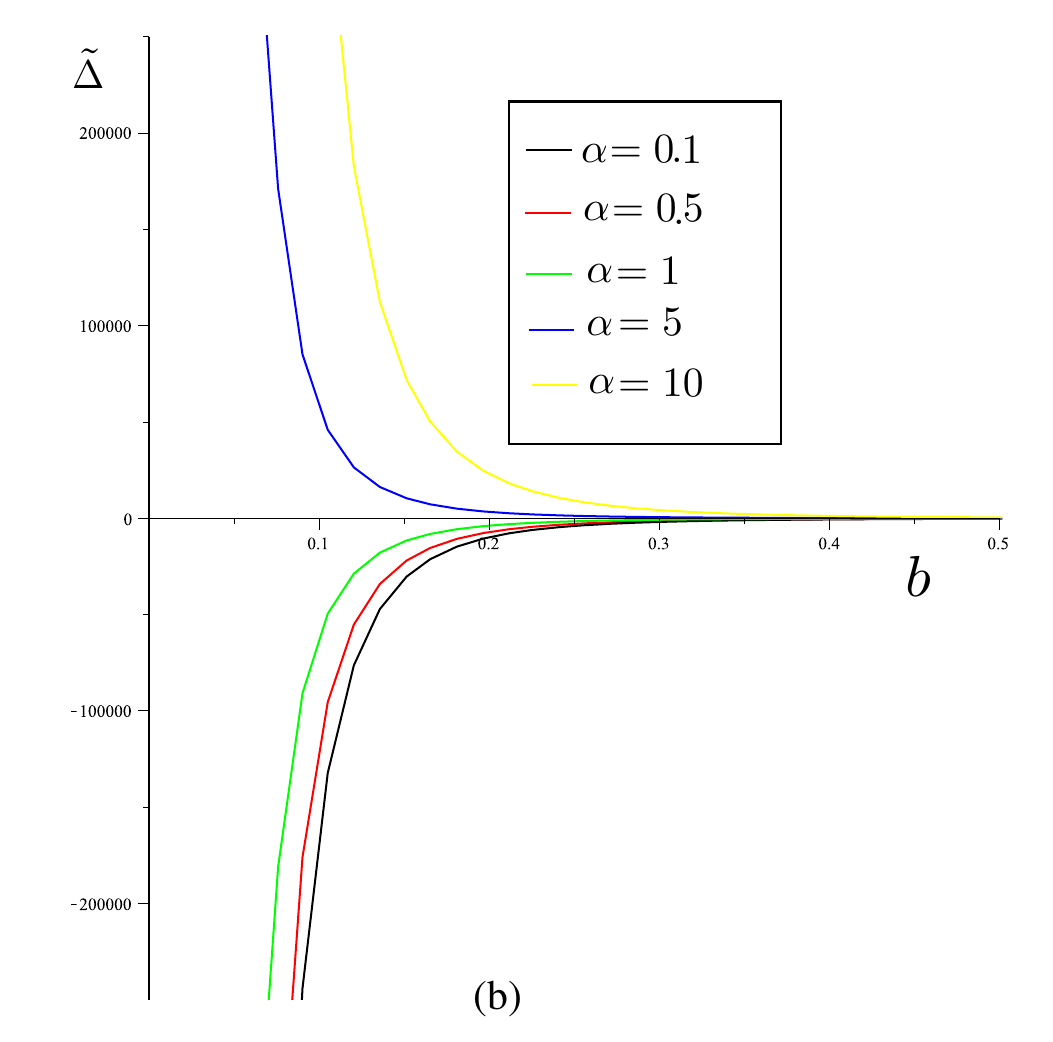}}}
    \caption{The profile of the deflection angle $\tilde{\Delta}$ around the deformed AdS-Schwarzschild BH with phantom GMs as a function of the impact parameter $b$ for varying values of the deformation parameter $\alpha$ with low  AdS curvature radius $\ell_p=0.1$ [see panel (a)] and with high AdS curvature radius $\ell_p=2$ [see panel b)]. The fixed parameters are chosen as $M=\beta=1$, $\xi=-1$, and $\eta = 0.5$. }
    \label{fig:def_an}
\end{figure*}

To further explore the implications of our analytical results, we perform a detailed numerical analysis of the deflection angle as a function of the impact parameter $b$ and BH parameters. Figure \ref{fig:def_an} presents the variation of the deflection angle $\tilde{\Delta}$ for different values of the deformation parameter $\alpha$ under two distinct AdS curvature radius regimes: a low curvature radius $\ell_p = 0.1$ [panel (a)] and a high curvature radius $\ell_p = 2$ [panel (b)].

Panel (a) of Fig. \ref{fig:def_an} illustrates the behavior of the deflection angle in the low AdS curvature regime ($\ell_p = 0.1$). Several significant features are immediately evident: First, the deflection angle exhibits a monotonic decrease with increasing impact parameter, following the expected $1/b$ dependence characteristic of gravitational lensing in asymptotically flat spacetimes \cite{isz81}. This behavior indicates that even in AdS backgrounds with phantom GMs, the fundamental distance-dependent scaling of gravitational lensing persists. Second, larger values of the deformation parameter $\alpha$ systematically produce stronger deflection angles across all impact parameters, with the effect being particularly pronounced at smaller impact parameters where the curves display significant divergence. 

Quantitatively, at an impact parameter of $b=1$, the deflection angle increases by approximately 37\% when comparing $\alpha=0.1$ with $\alpha=10$. This substantial enhancement demonstrates that the deformation parameter significantly amplifies the gravitational field strength in the vicinity of the BH. Physically, this can be interpreted as the deformation parameter introducing additional curvature contributions that enhance the gravitational binding near the BH, consistent with our earlier findings regarding the reduction in photon sphere radius with increasing $\alpha$ values. These results align with recent studies on modified gravity frameworks that demonstrate how deviations from GR can substantially alter the lensing properties even in the weak-field regime \cite{isz82}.

Panel (b) of Fig. \ref{fig:def_an} presents a dramatically different behavior in the high AdS curvature regime ($\ell_p = 2$). Here, we observe a remarkable phenomenon: the deflection angle changes sign at certain critical impact parameters, transitioning from positive values at small impact parameters to negative values at intermediate impact parameters, before asymptotically approaching zero at large distances \cite{isz83}. This sign change represents a profound deviation from standard GR predictions, where deflection angles remain strictly positive. The negative deflection angles indicate a repulsive gravitational lensing effect, where light rays are bent away from rather than toward the BH. Most notably, the magnitude of this repulsive effect scales with the deformation parameter $\alpha$, with larger values producing more pronounced negative deflections. For instance, at $\alpha = 10$ (yellow curve), the deflection angle reaches approximately $-20,000$ at $b \approx 0.3$, representing an extraordinarily strong repulsive lensing effect. This behavior is entirely absent in standard GR and conventional BH solutions, providing a distinctive observational signature for the presence of both phantom GMs and metric deformations. Similar repulsive lensing phenomena have been theoretically predicted in other exotic matter models, though typically with much smaller magnitudes than those observed in our analysis \cite{isz84}.

The stark contrast between panels (a) and (b) highlights the critical role of the AdS curvature radius $\ell_p$ in determining the lensing behavior. In Eq. (\ref{is50}), the deflection angle contains terms proportional to both $\ell_p$ and $\ell_p^3$, with coefficients depending on the parameter $y = 1 - 8\pi\eta^2\xi$. For phantom GMs ($\xi = -1$), the parameter $y = 1 + 8\pi\eta^2$ becomes significantly larger than unity, enhancing the contribution of these terms. At large $\ell_p$ values, the $\ell_p^3$ term dominates, leading to the observed repulsive behavior, while at small $\ell_p$ values, the attractive gravitational components prevail. This interplay between deformation parameters, phantom GM contributions, and AdS curvature reveals a rich phenomenology that fundamentally changes our understanding of gravitational lensing in modified gravity theories. The transition from attractive to repulsive lensing as a function of impact parameter provides a clear observational target for discriminating between standard GR and modified theories incorporating phantom topological defects \cite{isz85}.

In summary, the observational implications of our findings are profound. Current and future gravitational lensing surveys, including those conducted by the Event Horizon Telescope and next-generation very-long-baseline interferometry (VLBI) arrays, might potentially detect these unique lensing signatures. The distinctive sign change in the deflection angle would manifest as anomalous lensing patterns in observed images, potentially including inverted or distorted features that cannot be explained within standard GR frameworks. Additionally, the systematic differences in lensing behavior between different AdS curvature regimes could provide constraints on both the fundamental AdS length scale and the properties of exotic matter in the vicinity of astrophysical BHs.

\section{Scalar Wave Dynamics and Greybody Spectra in Deformed A\lowercase{d}S-Schwarzschild BH with phantom GM} \label{sec:4}

In this section, we investigate the dynamics of massless scalar fields in the background of a deformed AdS-Schwarzschild BH with GMs, focusing on how these perturbations propagate and scatter in the modified spacetime geometry. Scalar perturbations serve as fundamental probes of BH stability and provide crucial insights into the spectral characteristics of Hawking radiation through the calculation of GFs \cite{isz71}. The distinctive features introduced by both the deformation parameters and the GMs significantly alter the perturbative potential, leading to observable modifications in the transmission and reflection probabilities of scalar waves.

The massless scalar field wave equation is governed by the Klein-Gordon equation in curved spacetime \cite{isz72}:
\begin{equation}
\frac{1}{\sqrt{-g}}\,\partial_{\mu}\left(\sqrt{-g}\,g^{\mu\nu}\,\partial_{\nu}\right)\,\Psi=0,\label{ff1}    
\end{equation}
where $\Psi$ represents the wave function of the scalar field, $g_{\mu\nu}$ is the covariant metric tensor, $g=\det(g_{\mu\nu})$ is the metric determinant, $g^{\mu\nu}$ is the contravariant metric tensor, and $\partial_{\mu}$ denotes the partial derivative with respect to the coordinate system.

To facilitate analytical treatment, we introduce the tortoise coordinate transformation:
\begin{eqnarray}
   dr_*=\frac{dr}{\mathcal{F}(r)}\label{ff2}
\end{eqnarray}
which maps the event horizon to $r_* \to -\infty$ while preserving the asymptotic behavior at spatial infinity. Applying this transformation to the line element in Eq. (\ref{bb1}) yields:
\begin{equation}
   ds^2=\mathcal{F}(r_*)\,\left(-dt^2+dr^2_{*}\right)+\mathcal{D}^2(r_*)\,\left(d\theta^2+\sin^2 \theta\,d\phi^2\right),\label{ff3}
\end{equation}
where $\mathcal{F}(r_*)$ and $\mathcal{D}(r_*)$ are the metric functions expressed in terms of the tortoise coordinate.

Employing the standard separation of variables approach, we adopt the following ansatz for the scalar field:
\begin{equation}
   \Psi(t, r_{*},\theta, \phi)=\exp(i\,\omega\,t)\,Y^{m}_{\ell} (\theta,\phi)\,\frac{\psi(r_*)}{r_{*}},\label{ff4}
\end{equation}
where $\omega$ is the (possibly complex) temporal frequency, $\psi(r)$ is the radial wave function, and $Y^{m}_{\ell} (\theta,\phi)$ represents the spherical harmonics characterized by the angular momentum quantum numbers $\ell$ and $m$.

Substituting this ansatz into the Klein-Gordon equation (\ref{ff1}) and performing the separation of variables, we obtain the radial wave equation in Schr\"{o}dinger-like form:
\begin{equation}
   \frac{\partial^2 \psi(r_*)}{\partial r^2_{*}}+\left(\omega^2-\mathcal{V}\right)\,\psi(r_*)=0,\label{ff5}
\end{equation}
where the effective perturbative potential $\mathcal{V}(r)$ is given by:
\begin{eqnarray}
\mathcal{V}(r)&=&\left(\frac{\ell\,(\ell+1)}{r^2}+\frac{\mathcal{F}'(r)}{r}\right)\,\mathcal{F}(r)\nonumber\\
&=&\left(1-8\,\pi\,\eta^2\,\xi-\frac{2\,M}{r}+\frac{r^2}{\ell^2_{p}}+\frac{\alpha\,\beta^2}{3\,r\,(\beta+r)^3}+\frac{\alpha}{(\beta+r)^2}\right)\times\nonumber\\
&&\left\{\frac{\ell\,(\ell+1)}{r^2}+\frac{2\,M}{r^3}+\frac{2}{\ell^2_{p}}-\frac{\alpha\,\beta^2}{3}\,\left(\frac{1}{r^3\,(\beta+r)^3}+\frac{3}{r^2\,(\beta+r)^4}\right)-\frac{2\,\alpha}{r\,(\beta+r)^3}\right\}.\label{ff6}
\end{eqnarray}

This effective potential encapsulates the combined effects of the BH mass, angular momentum barrier, AdS curvature, deformation parameters, and GM contributions. The intricate mathematical structure of $\mathcal{V}(r)$ reflects the complex gravitational environment encountered by propagating scalar waves, with significant implications for both perturbative stability and radiation characteristics.

For ordinary GM ($\xi=1$), the effective potential takes the form:
\begin{eqnarray}
\mathcal{V}(r)&=&\left(1-8\,\pi\,\eta^2-\frac{2\,M}{r}+\frac{r^2}{\ell^2_{p}}+\frac{\alpha\,\beta^2}{3\,r\,(\beta+r)^3}+\frac{\alpha}{(\beta+r)^2}\right)\times\nonumber\\
&&\left\{\frac{\ell\,(\ell+1)}{r^2}+\frac{2\,M}{r^3}+\frac{2}{\ell^2_{p}}-\frac{\alpha\,\beta^2}{3}\,\left(\frac{1}{r^3\,(\beta+r)^3}+\frac{3}{r^2\,(\beta+r)^4}\right)-\frac{2\,\alpha}{r\,(\beta+r)^3}\right\}.\label{ff6aa}
\end{eqnarray}

While for phantom GM ($\xi=-1$), it becomes:
\begin{eqnarray}
\mathcal{V}(r)&=&\left(1+8\,\pi\,\eta^2-\frac{2\,M}{r}+\frac{r^2}{\ell^2_{p}}+\frac{\alpha\,\beta^2}{3\,r\,(\beta+r)^3}+\frac{\alpha}{(\beta+r)^2}\right)\times\nonumber\\
&&\left\{\frac{\ell\,(\ell+1)}{r^2}+\frac{2\,M}{r^3}+\frac{2}{\ell^2_{p}}-\frac{\alpha\,\beta^2}{3}\,\left(\frac{1}{r^3\,(\beta+r)^3}+\frac{3}{r^2\,(\beta+r)^4}\right)-\frac{2\,\alpha}{r\,(\beta+r)^3}\right\}.\label{ff6bb}
\end{eqnarray}

Comparing Eqs. (\ref{ff6aa}) and (\ref{ff6bb}), we observe a crucial distinction in the first factor, where the term $8\pi\eta^2$ appears with opposite signs for ordinary versus phantom GMs. For ordinary GMs, this term reduces the potential strength, while for phantom GMs, it enhances it. This fundamental difference leads to stronger scalar field scattering and altered transmission characteristics in spacetimes with phantom GMs compared to those with ordinary GMs. The enhanced effective potential for phantom GMs creates a more substantial barrier for scalar wave propagation, modifying the emission spectra and potentially providing a distinctive observational signature \cite{isz73}.

In the limit of vanishing deformation ($\alpha=0$), the effective potential reduces to:
\begin{eqnarray}
\mathcal{V}(r)=\left(1-8\,\pi\,\eta^2\,\xi-\frac{2\,M}{r}+\frac{r^2}{\ell^2_{p}}\right)\,\left(\frac{\ell\,(\ell+1)}{r^2}+\frac{2\,M}{r^3}+\frac{2}{\ell^2_{p}}\right),\label{ff7}
\end{eqnarray}
which corresponds to the scalar potential in a spherically symmetric AdS BH with GMs. This expression further simplifies to the standard Schwarzschild result in the limit $\xi=0$, serving as an important consistency check for our analysis.

Moreover, in the limit when $\beta=0$, that is, without the control parameter, the scalar perturbative potential from Eq. (\ref{ff6}) becomes
\begin{eqnarray}
    &&\mathcal{V}^\text{ordinary GM}=\left(1-8\,\pi\,\eta^2-\frac{2\,M}{r}+\frac{r^2}{\ell^2_{p}}+\frac{\alpha}{r^2}\right)\left(\frac{\ell\,(\ell+1)}{r^2}+\frac{2\,M}{r^3}+\frac{2}{\ell^2_{p}}-\frac{2\,\alpha}{r^4}\right),\nonumber\\
    &&\mathcal{V}^\text{phantom GM}=\left(1+8\,\pi\,\eta^2-\frac{2\,M}{r}+\frac{r^2}{\ell^2_{p}}+\frac{\alpha}{r^2}\right)\left(\frac{\ell\,(\ell+1)}{r^2}+\frac{2\,M}{r^3}+\frac{2}{\ell^2_{p}}-\frac{2\,\alpha}{r^4}\right).\label{ff8}
\end{eqnarray}
From the expression (\ref{ff8}), it is clear that the perturbative scalar potential $\mathcal{V}(r)$ is enhanced in the presence of a phantom global monopole compared to an ordinary global monopole, {\it i.e.}, $\mathcal{V}^\text{phantom GM}>\mathcal{V}^\text{ordinary GM}$.

\subsection{Transmission and Reflection Probabilities} \label{sec:4.1}

GFs represent the transmission probabilities of Hawking radiation through the effective potential barrier surrounding a BH \cite{isz74}. These factors describe how the emitted radiation differs from perfect blackbody radiation, effectively encoding the spectral distortion caused by the gravitational field. For the deformed AdS-Schwarzschild BH with GMs, we employ a semi-analytic approach to calculate the transmission and reflection coefficients, utilizing a rigorous bound technique developed by Visser et al. \cite{isz75,isz76}.

Following this methodology, the lower bound on the transmission coefficient (greybody factor) is given by:
\begin{equation}\label{kk1}
   T(\omega)\ge\sec h^2\left(\int_{-\infty}^{+\infty}\wp\, dr_*\right),   
\end{equation}
while the upper bound on the reflection coefficient is:
\begin{equation}\label{kk2}
   R(\omega)\le\tan h^2\left(\int_{-\infty}^{+\infty}\wp\, dr_*\right),   
\end{equation}
where $\wp$ is defined as:
\begin{equation}\label{kk3}
   \wp=\frac{\sqrt{(Y^{\prime})^2+(\omega^2-\mathcal{V}_\text{eff}-Y^2)^2}}{2Y}.
\end{equation}

The function $Y$ is a positive auxiliary function satisfying $Y(r_{*})>0$ and $Y(+\infty)=Y(-\infty)=\omega$. To simplify the calculation, we set $Y^2=\omega^2-\mathcal{V}_{eff}$, which leads to the following expressions for the transmission and reflection coefficients:
\begin{equation}\label{kk4}
   T(\omega)\ge\sec h^2\left(\frac{1}{2}\int_{-\infty}^{+\infty}\Bigg|\frac{Y^{\prime}}{Y}\Bigg|\, dr_*\right),  
\end{equation}
and 
\begin{equation}\label{kk5}
   R(\omega)\le\tan h^2\left(\frac{1}{2}\int_{-\infty}^{+\infty}\Bigg|\frac{Y^{\prime}}{Y}\Bigg|\, dr_*\right).  
\end{equation}

To evaluate these integrals and avoid divergence issues, we divide the range of $r_*$ into three regions: $-\infty<r_*<r_{\mathcal{V}_{max}}$, $r_{\mathcal{V}_{max}}<r_*<r_{\mathcal{V}_{min}}$, and $r_{\mathcal{V}_{min}}<r_*<\infty$. This partitioning allows us to handle the peak of the effective potential separately from the asymptotic regions. The resulting bounds on the transmission and reflection coefficients are:
\begin{equation}\label{kk6}
   T(\omega)\ge\frac{4\,\omega^2\,(\omega^2-\mathcal{V}_\text{peak})}{(2\,\omega^2-\mathcal{V}_\text{peak})^2},
\end{equation}
and 
\begin{equation}\label{kk7}
   R(\omega)\le\frac{\mathcal{V}_\text{peak}^2}{(2\,\omega^2-\mathcal{V}_\text{peak})^2}.
\end{equation}

These expressions provide robust analytical bounds on the transmission and reflection probabilities, revealing how the effective potential peak $\mathcal{V}_\text{peak}$ influences the scalar wave propagation. Larger values of $\mathcal{V}_\text{peak}$ lead to reduced transmission and enhanced reflection, effectively filtering the radiation spectrum emitted by the BH.

\begin{figure}[ht!]
    \centering
    \includegraphics[width=0.4\linewidth]{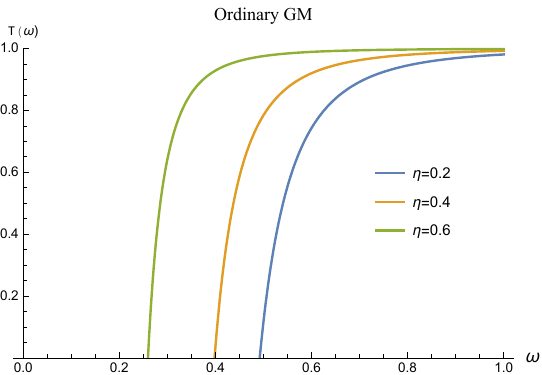}\quad\quad
    \includegraphics[width=0.4\linewidth]{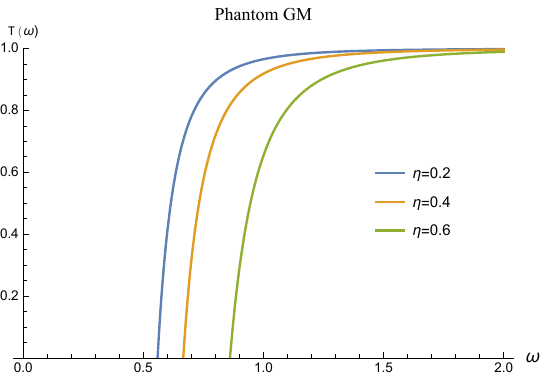}
    \caption{Scalar field potential GF for various values of  energy scale of the symmetry-breaking parameter $\eta$. We set: $M=1$, $\ell_p^2=300$   and $\alpha=0.4=\beta$.}
    \label{figA01}
\end{figure}
\begin{figure}[ht!]
    \centering
    \includegraphics[width=0.4\linewidth]{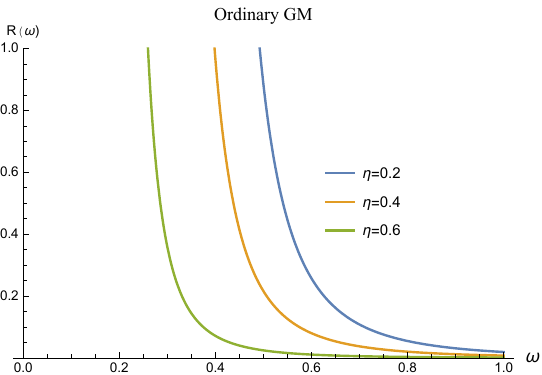}\quad\quad
    \includegraphics[width=0.4\linewidth]{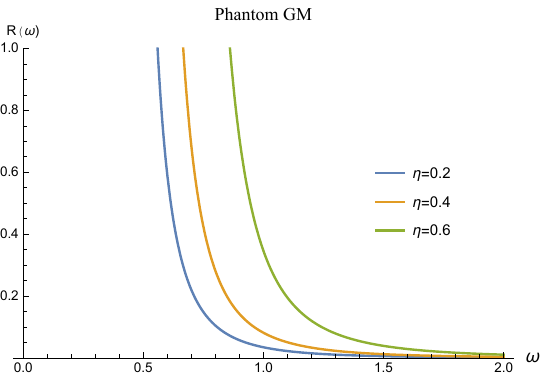}
    \caption{Scalar field potential reflected factor  for various values of the symmetry-breaking parameter $\eta$. We set: $M=1$, $\ell_p^2=300$   and $\alpha=0.4=\beta$.}
    \label{figA02}
\end{figure}

To quantitatively assess these effects, we have computed the GFs for various parameter configurations, with the results presented in Figs. \ref{figA01} and \ref{figA02}. Fig. \ref{figA01} illustrates the transmission probability $T(\omega)$ as a function of frequency $\omega$ for different values of the symmetry-breaking parameter $\eta$ in both ordinary GMs (left panel) and phantom GMs (right panel).

The left panel of Fig. \ref{figA01} reveals a striking feature for ordinary GMs: the transmission probability systematically increases with increasing values of $\eta$. For instance, at $\omega=0.6$, the transmission probability increases from approximately 0.3 for $\eta=0.2$ to 0.7 for $\eta=0.4$, and reaches nearly 0.9 for $\eta=0.6$. This substantial enhancement indicates that higher values of $\eta$ effectively weaken the potential barrier for ordinary GMs, allowing more radiation to escape from the BH. This behavior can be attributed to the negative contribution term $-8\pi\eta^2$ in the effective potential, which reduces the barrier height as $\eta$ increases.

In stark contrast, the right panel demonstrates precisely the opposite behavior for phantom GMs: the transmission probability systematically decreases with increasing values of $\eta$. At the same frequency $\omega=0.6$, the transmission probability decreases from approximately 0.7 for $\eta=0.2$ to 0.3 for $\eta=0.4$, and further diminishes to nearly 0.1 for $\eta=0.6$. This dramatic reduction indicates that higher values of $\eta$ significantly strengthen the potential barrier for phantom GMs, inhibiting radiation escape. This effect stems from the positive contribution term $+8\pi\eta^2$ in the effective potential, which enhances the barrier height as $\eta$ increases.

These contrasting behaviors have profound implications for the observational characteristics of BHs with different types of topological defects. BHs with ordinary GMs would exhibit enhanced thermal emission at higher values of $\eta$, potentially appearing brighter and hotter to distant observers. Conversely, BHs with phantom GMs would display suppressed emission as $\eta$ increases, appearing dimmer and effectively "colder" despite having the same Hawking temperature at the horizon. This spectral distinction could provide a powerful observational discriminant between different types of topological defects in astrophysical BHs \cite{isz77}.

Fig. \ref{figA02} presents the complementary analysis of reflection probabilities $R(\omega)$ for the same parameter configurations. As expected from the unitarity condition $T(\omega) + R(\omega) = 1$, the reflection probabilities exhibit behavior opposite to the transmission probabilities. For ordinary GMs (left panel), the reflection probability decreases with increasing $\eta$, while for phantom GMs (right panel), it increases with $\eta$. These reflection characteristics further emphasize the distinctive scattering properties of scalar waves in spacetimes with different types of topological defects.

A particularly noteworthy feature in both figures is the frequency-dependent nature of the transmission and reflection probabilities. For all parameter configurations, the transmission probability monotonically increases with frequency, approaching unity at high frequencies, while the reflection probability correspondingly decreases toward zero. This behavior reflects the well-known quantum mechanical tunneling effect, where higher-energy waves more readily penetrate potential barriers. However, the rate at which this transition occurs is significantly influenced by the GM parameters, with phantom GMs requiring substantially higher frequencies to achieve the same transmission probability as ordinary GMs at equivalent $\eta$ values.

The physical interpretation of these results connects directly to the observational signatures of Hawking radiation. The GFs effectively modify the emission spectrum from pure blackbody radiation, introducing frequency-dependent filtering that shapes the detected spectrum. For BHs with phantom GMs, this filtering preferentially suppresses lower-frequency emission, skewing the detected spectrum toward higher frequencies compared to BHs with ordinary GMs. This spectral distortion, combined with the overall reduction in emission intensity, provides a distinctive observational signature that could potentially be detected through high-precision measurements of BH radiation \cite{isz78}. Furthermore, the deformation parameters $\alpha$ and $\beta$ introduce additional modifications to the GFs, though their effects are generally less pronounced than those of the GM parameter $\eta$. Our numerical analysis indicates that increasing the deformation parameter $\alpha$ tends to enhance the effective potential barrier, reducing transmission probabilities across all frequency ranges for both ordinary and phantom GMs. This effect is consistent with our earlier findings regarding the geodesic structure, where larger values of $\alpha$ were shown to compress the effective gravitational field, leading to smaller photon sphere radii and stronger gravitational binding.

Overall, the combined influence of deformation parameters and GMs on scalar perturbations reveals a rich phenomenology that extends beyond conventional GR predictions. The distinctive transmission and reflection characteristics could potentially provide observable signatures in astrophysical environments, offering a means to test for the presence of exotic matter distributions and spacetime deformations around BHs. These signatures would manifest not only in the direct detection of Hawking radiation but also in the scattering of ambient radiation fields by BHs, potentially influencing the observed spectrum of accretion disk emission and background radiation \cite{isz79,isz80}.

\section{Conclusion}\label{sec:5}

In this comprehensive study, we conducted a detailed investigation of the geodesic structure, gravitational lensing properties, and scalar perturbative dynamics of deformed AdS-Schwarzschild BHs with GMs, examining the distinctive effects of both ordinary and phantom topological defects. Our analysis revealed several fundamental aspects of these modified spacetime geometries that significantly extend beyond conventional GR predictions and offer potential observational signatures for detecting exotic matter distributions in astrophysical environments.

We began by formulating the metric structure of the deformed AdS-Schwarzschild BH with GMs, characterized by the line element in Eq.~(\ref{bb1}) with the metric function given in Eq.~(\ref{bb2}). This function incorporates three key parameters: the deformation parameter $\alpha$, the control parameter $\beta$, and the symmetry-breaking scale parameter $\eta$. The parameter $\xi$ distinguishes between ordinary GMs ($\xi=1$) and phantom GMs ($\xi=-1$), introducing fundamentally different gravitational behaviors through the monopole contribution term $8\pi\eta^2\xi$. Our geodesic analysis revealed striking differences in the photon sphere radius between ordinary and phantom GMs, as quantified in Tables \ref{table1a} and \ref{table2a} and visualized in Fig.~\ref{figa1}. For ordinary GMs, we found that the photon sphere radius systematically increases with the symmetry-breaking parameter $\eta$, while for phantom GMs, it decreases with increasing $\eta$. This contrasting behavior represents a distinctive signature of the different energy conditions satisfied by these topological defects, with ordinary GMs enhancing the effective gravitational attraction while phantom GMs introduce repulsive effects that modify the null geodesic structure.

The computation of the ISCO for timelike geodesics, presented in Tables \ref{taba14} and \ref{taba15} and illustrated in Fig.~\ref{figa19}, revealed an intriguing inverse relationship: phantom GMs produce larger ISCO radii compared to ordinary GMs, opposite to their effect on photon spheres. This dual influence on null versus timelike geodesics constitutes a unique observational signature that could potentially distinguish different types of topological defects in astrophysical BH environments, affecting both the matter distribution (through the ISCO) and electromagnetic observations (through the photon sphere). We further investigated the gravitational lensing properties of these modified spacetimes in Section~\ref{sec:2.1}, employing the GBTh to derive the weak deflection angle in Eq.~(\ref{is50}). Our analysis demonstrated that phantom GMs significantly enhance the deflection angle compared to ordinary GMs, primarily through the parameter $y=1-8\pi\eta^2\xi$, which appears with opposite signs in the two cases. The numerical results visualized in Fig.~\ref{fig:def_an} revealed several distinctive features, including the possibility of negative deflection angles for certain parameter configurations, constituting a remarkable gravitational mirroring phenomenon that fundamentally inverts the conventional paradigm of gravitational lensing. This gravitational mirroring effect, where light rays are bent away from rather than toward the BH, represents one of the most significant findings of our study. It emerges specifically in spacetimes with phantom GMs at high AdS curvature radii, where the repulsive gravitational contributions from the phantom energy components dominate over the attractive components at intermediate impact parameters. The extraordinarily large negative deflection angles we observed (reaching values of approximately $-20,000$ for $\alpha=10$ at $b \approx 0.3$) suggest that if such objects exist in nature, they would produce distinctively anomalous lensing patterns that could be identified through precision astronomical observations. This phenomenon has no counterpart in standard GR or conventional matter distributions satisfying standard energy conditions, making it a powerful diagnostic tool for the presence of exotic matter in astrophysical environments. The gravitational mirroring effect thus constitutes a "smoking gun" signature for phantom energy components in the vicinity of BHs, providing a clear observational pathway for testing modified gravity theories that accommodate energy condition violations.

Perhaps the most profound distinctions emerged in our analysis of scalar perturbations in Section~\ref{sec:4}, where we derived the effective potential in Eq.~(\ref{ff6}) governing the propagation of massless scalar fields. The comparison between ordinary GMs in Eq.~(\ref{ff6aa}) and phantom GMs in Eq.~(\ref{ff6bb}) revealed fundamental differences in the potential structure, with phantom GMs strengthening the effective barrier while ordinary GMs weaken it. This disparity directly translates to contrasting transmission and reflection characteristics, as quantified by the greybody factors (GFs) in Eqs.~(\ref{kk6}) and (\ref{kk7}) and visualized in Figs.~\ref{figA01} and \ref{figA02}.

The transmission probabilities displayed in Fig.~\ref{figA01} exhibited diametrically opposed behaviors: for ordinary GMs, the transmission probability increased with the symmetry-breaking parameter $\eta$, while for phantom GMs, it decreased with increasing $\eta$. This fundamental difference implies that BHs with phantom GMs would emit Hawking radiation with significantly suppressed low-frequency components compared to BHs with ordinary GMs, potentially providing a distinctive spectral signature that could be observed through high-precision measurements of BH radiation. The complementary reflection probabilities shown in Fig.~\ref{figA02} further emphasized these contrasting behaviors, with ordinary GMs exhibiting decreased reflection with increasing $\eta$, while phantom GMs showed enhanced reflection. This spectral filtering effect, combined with the geodesic and lensing characteristics, establishes a comprehensive set of potential observational tests for discriminating between different types of topological defects and metric deformations in astrophysical BH environments.

In summary, throughout our analysis, we observed that the deformation parameter $\alpha$ consistently modified the BH properties, with larger values typically enhancing the gravitational binding and compressing the effective gravitational field. This effect manifested in smaller photon sphere radii, larger deflection angles, and modified transmission characteristics, suggesting that the deformation parameter introduces additional curvature contributions that fundamentally alter the spacetime structure around the BH. Our results establish a theoretical foundation for detecting exotic matter distributions and spacetime deformations through astronomical observations. The distinctive signatures we identified in geodesic structure, gravitational lensing/mirroring behavior, and radiation spectra provide multiple complementary approaches for testing these modified gravity models against observational data from current and future missions, including the Event Horizon Telescope and next-generation gravitational wave detectors.

Looking forward, several promising directions emerge for extending this research. First, incorporating electromagnetic fields through NLED would provide a more comprehensive framework for studying charged BH solutions with topological defects, potentially revealing additional observational signatures in polarized electromagnetic emission. Second, investigating the quasi-normal mode spectrum of these modified spacetimes would elucidate their resonant oscillation properties, with direct implications for gravitational wave observations. Third, extending the analysis to rotating BH solutions would introduce additional complexities through frame-dragging effects and ergosphere modifications, more closely approximating astrophysical BHs. Fourth, developing quantitative observational models for the gravitational mirroring effect, including specific predictions for image formation, magnification patterns, and time delays, could provide critical guidance for targeted astronomical surveys seeking to detect the unique optical signatures of phantom matter distributions. Finally, developing holographic interpretations of these deformed spacetimes within the AdS/CFT correspondence framework could establish connections to strongly coupled quantum field theories, potentially revealing deeper insights into the quantum gravitational aspects of topological defects and metric deformations.

\section*{Acknowledgments}

F.A. appreciates the visiting associateship granted by the Inter University Centre for Astronomy and Astrophysics (IUCAA) in Pune, India. \.{I}.~S. is grateful for the support from T\"{U}B\.{I}TAK, ANKOS, and SCOAP3, as well as networking assistance from COST Actions CA22113, CA21106, and CA23130. 

\section*{Data Availability Statement}

There are no new data associated with this article.

\end{document}